\documentclass[preprint,prfluids,superscriptaddress,eqsecnum]{revtex4-2} 

\usepackage{natbib}
\usepackage{graphicx}
\usepackage{subfigure}
\usepackage{amssymb}
\usepackage{amsfonts}

\usepackage{amsmath}
\usepackage{amsbsy}
\usepackage{mathrsfs} 
\usepackage{color}
\usepackage[colorlinks=true,citecolor=blue]{hyperref} 
\usepackage{soul}
\usepackage{lineno}
\usepackage{mymacros}
\usepackage{graphicx}
\begin{document} 
\begin{abstract}
We calculate the effective slip length for a rectangularly grooved periodic surface encapsulated (i.e., fully wetted) by a lubricant fluid and subjected to exterior shear flow parallel to the grooves. Our focus is the limit of a nearly-inviscid lubricant, where the ratio $\mu$ of the lubricant viscosity to that of the exterior fluid is small. This limit is singular for an encapsulated surface, indicating a dominant lubricant-flow effect --- a stark contrast to superhydrophobic surfaces where the role of the lubricant is typically negligible. In addition to $\mu$, the ratio $\lambda$ of the slip length to the grooving semi-period depends on three geometric lengths, all normalized by the semi-period: $b$, the thickness of the lubricant films wetting the groove ridges, and $\phi$ and $h$, the semi-width and height of the ridges, respectively. We identify two key limits characterizing the regime $\mu\ll1$. In the first, with $b$ held fixed, we find $\lambda\sim\mu^{-1}\tilde{\lambda}(b,\phi,h)$, where the rescaled slip length $\tilde{\lambda}$ is determined by an interior lubricant-flow problem. In the second, with $b/\mu$ held fixed, we find $\lambda\sim\Lambda(b/\mu,\phi)$, where $\Lambda$ is governed by an exterior flow problem in which the thin lubricant films wetting the ridges are effectively replaced by a Navier-slip condition, while the rest of the interface is shear-free. As $b/\mu\to0$, the Navier-slip condition simplifies to no slip, whereby the exterior problem reduces to that for a superhydrophobic grooved surface --- famously solved by Philip using complex variables (Z.~Angew.~Math.~Phys., 23 353, 1972). Asymptotic and numerical analysis of the exterior problem demonstrates (i) the transition from Philip's solution at $b/\mu\ll1$ to an algebraic behavior $\Lambda\sim b/(\mu\phi)$ at $b/\mu\gg1$, matching with the small-$b$ limit of the interior problem; and (ii) for $\phi\ll1$, the transition from a logarithmic $O(\ln\phi)$ scaling for $b\ll\mu\phi$ --- familiar from the superhydrophobic case, where $\phi$ constitutes the solid-to-air fraction --- to the aforementioned algebraic regime when $\mu\phi\ll b\ll\phi$. As $b$ becomes comparable to $\phi$, the exterior problem loses validity in favor of the interior problem. By analyzing the latter in the distinguished sub-limit where $b$ and $\phi$ are comparably small, we demonstrate how, as $b/\phi$ is increased, the algebraic growth of $\lambda$ with $b/\mu$ is arrested at order $\mu^{-1}/\ln b$. 
\end{abstract}

\title{Effective longitudinal slip over grooves \\ encapsulated by a nearly inviscid lubricant}
\author{Ory Schnitzer}
\affiliation{Department of Mathematics, Imperial College London, London SW7 2AZ, UK}
\author{Ehud Yariv}
\affiliation{Department of Mathematics, Technion---Israel Institute of Technology, Haifa 32000, Israel}
\maketitle

\section{Introduction}
Microstructured surfaces designed to entrap a lubricant fluid offer potential for reducing hydrodynamic resistance on small scales. Traditionally, research has focused on superhydrophobic surfaces, which combine hydrophobic materials and geometric roughness to trap air pockets within the cavities of the microstructure \citep{Quere:08,Rothstein:10,Bocquet:11}. In recent years, slippery liquid-infused porous surfaces (SLIPS), where a liquid lubricant partially or fully wets the microstructure, have emerged as a promising alternative offering enhanced stability and durability \citep{Lafuma:11,Wong:11,Guan:17,Keiser:17,Kreder:18,Sadullah:18,Villegas:19,Keiser:20,Nath:24}. Although the primary function of the lubricant is often to prevent the pinning of contact lines, 
liquid infusion simultaneously offers a mechanism to reduce the hydrodynamic drag experienced by bulk flows \citep{Solomon:14,Hardt:22}.


Drag reduction is often theoretically characterized using a hydrodynamic effective-slip problem where an infinite compound surface is subjected to a simple shear flow \citep{Davis:09,Crowdy:10:slip}. In this framework, the flow enhancement due to the microstructure is quantified by the effective slip length, a key integral output of the problem defined as the apparent displacement of the shear flow ``below'' the surface in a far-field expansion of the velocity field (i.e., at large distances from the surface). At small scales and low speeds, inertia and interfacial deformation are typically negligible, rendering the effective-slip problem linear. In that regime, the dimensionless slip length (normalized by a characteristic length scale, say the microstructure semi-period) depends on the geometry of the microstructure and fluid interfaces; the viscosity ratio $\mu$ between the lubricant and the exterior liquid; and the direction of the applied shear flow in the plane parallel to the surface. Owing to this linearity, it suffices to consider shear flow along, at most, two principal directions. 

Until recently, analyses of the effective-slip problem have focused on superhydrophobic surfaces, where the minute viscosity of the trapped air relative to the exterior (``working'') fluid dictates an exceedingly small $\mu$ value \citep{Lee:16}. Based on this, most analyses have assumed that the exterior fluid satisfies a shear-free condition at its interface with the trapped air. Under that simplification, the lubricant flow is eliminated from the problem; the resulting exterior ``inviscid'' problem is that of (viscous) exterior liquid flow over a compound surface composed of no-slip (solid) and shear-free (free surface) regions.

A prototypical superhydrophobic surface consists of a periodically grooved hydrophobic solid substrate sustaining a ``Cassie'' wetting state, in which air spontaneously fills the grooves. Assuming flat menisci pinned to the groove edges, the inviscid problem considers a flat boundary composed of alternating no-slip and shear-free strips. In this canonical inviscid problem, the geometry is defined solely by the solid fraction of the boundary; the depth of the grooves does not affect the exterior-fluid flow. Furthermore, owing to symmetry, the principal shear directions are parallel and perpendicular to the grooves, respectively defining ``longitudinal'' and ``transverse'' slip lengths. Philip \cite{Philip:72,Philip:72:integral} famously solved both the longitudinal and transverse cases using complex-variable methods, though motivated by porous-media flows rather than superhydrophobic surfaces. Subsequent studies have extended the canonical inviscid problem to account for variations such as non-flat menisci \citep{Haase:16,Crowdy:16,Crowdy:17,Schnitzer:17,Luca:18}, partially invaded grooves \citep{Ng:09,Crowdy:11,Crowdy:21,Yariv:23,Yariv:23:tr,Yariv:23:highly,Yariv:24:grating}, complex fluids \citep{Crowdy:17:Thinning,Schnitzer:24thinning}, and surfactant contamination \citep{Peaudecerf:17,Song:18,Baier:21,Landel:20,Sundin:22,Mayer:22,Baier:22,Rodriguez:23}. Additionally, related inviscid problems have been analyzed considering more involved configurations, including doubly periodic microstructures \cite{Davis:09:mesh,Davis:10,Schnitzer:18,Schnitzer:18:Fakir} and pressure-driven flow through superhydrophobic channels \citep{Lauga:03,Ou:04,Davies:06,Sbragaglia:07,Teo:09,Schnitzer:17:narrow,Yariv:17:amplification,Yariv:18}. 

Recently, there has been growing interest in understanding how effective slip is influenced by the lubricant flow \citep{Schonecker:14,Nizkaya:14}. The motivation for this is twofold. First, to assess the effect in the superhydrophobic scenario, where $\mu$ is naturally small and the lubricant effect is traditionally neglected. The second has to do with the emergence of SLIPS, where the lubricant is a viscous liquid rather than air; here, a relatively inviscid lubricant seems desirable for achieving significant drag reduction. In both cases, the limit of interest is accordingly $\mu\ll1$ (henceforth termed the ``nearly-inviscid'' limit). Intuitively, small $\mu$ suggests only a minor perturbation from the inviscid problem. Indeed, Crowdy \cite{Crowdy:17:Perturbation} calculated such small perturbations for grooved superhydrophobic surfaces using complex-variable methods and a regular expansion about Philip's solution of the corresponding inviscid problem. 

Can the intuition that small $\mu$ implies small lubricant-flow effects fail? Generally speaking, such failure implies a scenario where the nearly-inviscid limit $\mu\ll1$ becomes singular; i.e., where the inviscid problem --- corresponding to setting $\mu=0$ --- becomes ill-posed. 
One such scenario is that of zero solid-fraction, $\phi=0$, where $\phi$ represents the area fraction of the exterior-fluid boundary that is in contact with the solid substrate. Indeed, Philip's solution of the canonical inviscid problem gives a slip length that diverges logarithmically as $\phi\to0$. Scaling arguments and asymptotic analysis of the inviscid problem for various other configurations have been widely employed to characterize similar logarithmic singularities \citep{Ybert:07}, as well as stronger ``algebraic'' ones where the dimensionless slip length scales as some inverse power of $\phi$ \citep{Schnitzer:16,Yariv:23:highly}. In accordance with these predictions, very large slip lengths of small-solid-fraction superhydrophobic surfaces have indeed been observed \citep{Choi:06,Lee:08}, albeit under carefully maintained laboratory conditions. 

As a first step in clarifying the singular effect of  nearly-inviscid lubricant flow, Peng \textit{et al.}  \cite{Peng:25} recently studied the aforementioned scenario in the case of lubricant-filled grooves formed of zero-thickness ridges, with the exterior fluid in contact with the solid substrate only at the ridge tips. While this problem is well posed for any fixed $\mu$, the corresponding inviscid problem is ill-posed --- indeed, it corresponds to Philip's canonical problem at vanishing solid fraction. Hence, in this case the lubricant flow plays an essential role for arbitrarily small $\mu$. Focusing on the singular limit $\mu\to0$, Peng \textit{et al.} \cite{Peng:25} showed that the drag is dominated by exponentially small regions about the ridge tips where the lubricant and exterior-fluid flows are both important, leading to dimensionless slip lengths scaling as $\mu^{-1/2}$. 

In the present paper, we consider another scenario where a nearly inviscid lubricant dramatically influences effective slip: encapsulated SLIPS, where the liquid lubricant fully wets the solid substrate. As discussed in Ref.~\citep{Smith:13}, an encapsulated state implies a positivity condition on the spreading parameter measuring the tendency of the lubricant to wet the solid whilst displacing the exterior fluid. For encapsulated SLIPS, the inviscid problem is ill-posed regardless of the microstructure geometry. Indeed, with encapsulation normally implying a flat fluid interface, the exterior fluid satisfies a shear-free condition over that flat boundary. In that case, the imposed stress associated with the far-field shear flow cannot be supported \footnote{This is the same singularity approached in Philip's canonical inviscid problem as the solid fraction is made to vanish. Unlike with encapsulation, however, the small-solid-fraction limit of the inviscid problem is not universally ill-posed. An example for this is the transverse problem for a grooved superhydrophobic surface with protruding menisci.}.

As the inviscid problem for encapsulated SLIPS is ill-posed, we expect the slip length to diverge in the nearly-inviscid limit $\mu\to0$ --- but at what rate? We shall use asymptotic methods to analyze this limit in the case of longitudinal shear flow over a grooved substrate. As we shall see, the divergence rate is highly sensitive to the height of the encapsulation above the microstructure, as well as the geometry of the submerged microstructure. In particular, our analysis will mostly focus on regimes where the encapsulation height at the ridge tops is small relative to the period. 

\section{Problem formulation}
\begin{figure}[t!]
\begin{center}
\includegraphics[scale=0.8]{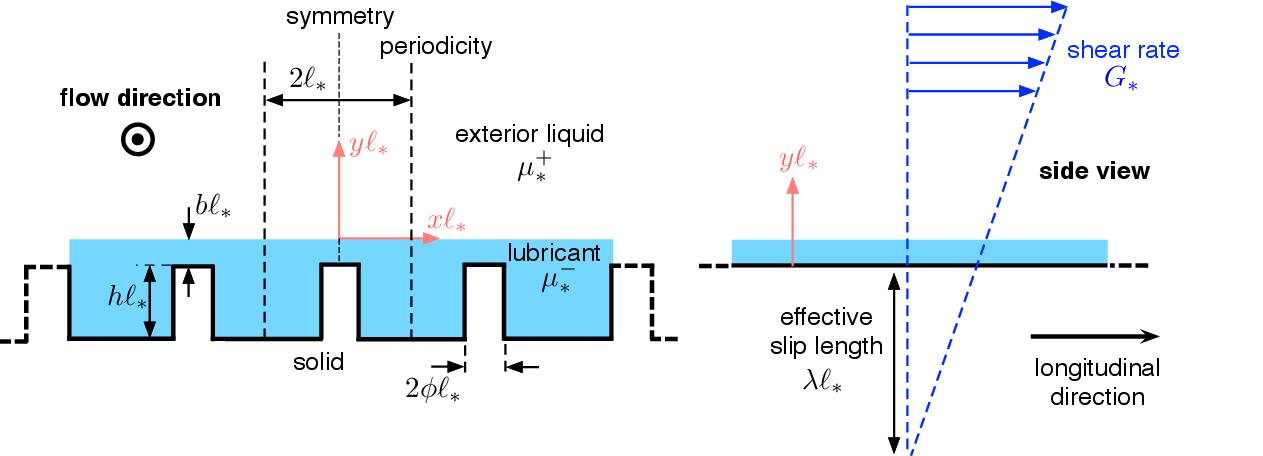}
\caption{Dimensional schematic.}
\label{fig:dimsketch}
\end{center}
\end{figure}

We study shear-driven flow over a microstructured solid substrate. The microstructure consists of a periodic array of rectangular grooves; the period of the groove pattern is denoted by $2\ell_*$, the width of the ridges separating the grooves by $2\phi\ell_*$ and their height (the groove depth) by $h\ell_*$.  (Throughout the paper, we use an asterisk subscript to indicate a dimensional quantity). The substrate is encapsulated by a lubricant liquid of viscosity $\mu_*^-$ and overlaid by an exterior liquid of viscosity $\mu_*^+$. The flat interface between the two liquids is  positioned at a distance $b\ell_*$ from the ridge tops. A simple shear flow of rate $G_*$ is applied far from the substrate in the longitudinal direction along the grooves. A schematic is shown in Fig.~\ref{fig:dimsketch}. 

\begin{figure}[t!]
\begin{center}
\includegraphics[scale=0.8]{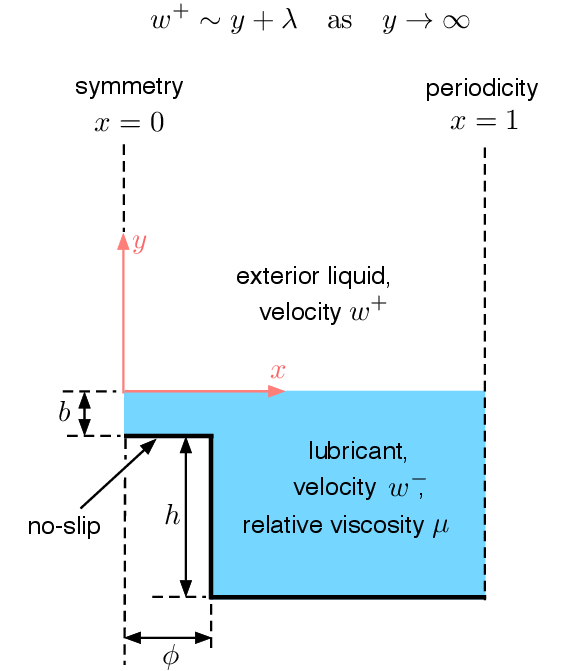}
\caption{Dimensionless schematic.}
\label{fig:dimless}
\end{center}
\end{figure}

Henceforth, we adopt a dimensionless convention where lengths are normalized by the half-period $\ell_*$ and velocities  by $G_*\ell_*$. We employ dimensionless Cartesian coordinates $(x,y)$ with origin at the interface point above the center of one of the ridge tops; the $x$ axis runs along the interface, perpendicular to the longitudinal direction, while the $y$ axis points away from the substrate. The velocity field is unidirectional, possessing a single component along the shear direction. For the exterior fluid, this component  is denoted by $w^+$, while for the lubricant liquid it is denoted by $w^-$. Besides cross-sectional position, these two fields depend on the viscosity ratio $\mu=\mu_*^-/\mu_*^+$ and the three geometric parameters: ridge width $\phi$, ridge height (i.e., groove depth) $h$ and encapsulation height $b$. The periodicity and symmetry of the problem allows us to restrict the problem formulation to one semi-period, say $x\in[0,1]$. The dimensionless geometry in that ``semi-unit-cell'' is depicted in Fig.~\ref{fig:dimless}.

The velocity fields $w^{\pm}$ satisfy Laplace's equation,
\begin{equation} \label{Laplace}
\pd{^2w^\pm}{x^2}+\pd{^2w^\pm}{y^2} = 0,
\end{equation}
 in their respective domains. At the solid boundary, the lubricant satisfies a no-slip condition, 
\begin{equation}\label{no slip}
w^-=0 ,
\end{equation}
which introduces a dependence upon $b$, $h$, and $\phi$.
At the liquid-liquid interface, $y=0$, continuity of velocity and tangential stress provides the conditions
\refstepcounter{equation}
$$
\label{interfacial}
w^-=w^+, \quad \mu \pd{w^-}{y} = \pd{w^+}{y},
\eqno{(\theequation{\mathit{a},\mathit{b}})}
$$
which introduce a dependence upon $\mu$.
Symmetry and periodicity imply the homogeneous Neumann conditions  
\begin{equation}\label{sym per}
\pd{w^{\pm}}{x}=0 \quad \text{at} \quad x=0,1. 
\end{equation}
Lastly, we have the far-field behavior 
\begin{equation}\label{far}
w^+ = y + \lambda + o(1) \quad \text{as} \quad y\to\infty.
\end{equation}
The leading linear term, which represents the imposed simple shear flow, is prescribed. The constant correction $\lambda$, which corresponds to the effective slip length normalized by the half-period, remains to be found as part of the solution; it is the output of interest. 

Following Schnitzer \cite{Schnitzer:16}, we obtain a useful relation by integrating over Laplace's equation in the exterior fluid. Making use of the two-dimensional variant of Gauss's theorem in conjunction with \eqref{sym per} and \eqref{far}, we thereby find 
\begin{equation}\label{int out}
\int_0^1\pd{w^+}{y}\,\dd x = 1 \quad \text{at} \quad y=0.
\end{equation}
This integral relation represents a balance between the far-field unity stress associated with the imposed simple shear flow and the interfacial shear stresses averaged over a semi-period. Alternatively, using \eqref{interfacial}, 
\begin{equation}\label{int in}
\mu\int_0^1\pd{w^-}{y}\,\dd x = 1 \quad \text{at} \quad y=0.
\end{equation}
Since the integral relations \eqref{int out}--\eqref{int in} follow from the problem formulation, they do not provide independent information. Nonetheless, they are indispensable for the scaling arguments and asymptotic analysis that follow. 

\section{Nearly-inviscid lubricant}\label{sec:nearly} 
Henceforth, our analysis of the problem posed by \eqref{Laplace}--\eqref{far} focuses on  the limit $\mu\to0$. In forming  this limit, we shall initially treat the geometric parameters $b$, $\phi$ and $h$ as fixed constants. As discussed in the Introduction, the encapsulation of the substrate implies that the limit $\mu\ll1$ is singular. Indeed, the problem is clearly ill-posed for $\mu=0$. We therefore anticipate that the slip length $\lambda$ diverges as $\mu\to0$.

The actual scaling of this divergence follows from \eqref{int in}, which suggests that $w^-$ is of order $\mu^{-1}$. Because of the continuity condition (\ref{interfacial}\textit{a}), this is also the scaling of $w^+$,
\begin{equation}
w^+ \sim \mu^{-1} \tilde w^+ \quad\text{as} \quad \mu\to0.
\end{equation}
The integral relation \eqref{int out} then suggests that $w^+$ is uniform at $\ord(\mu^{-1})$, say
\begin{equation}\label{w plus is tilde lambda}
\tilde w^+ \equiv \tilde\lambda.
\end{equation}
Given the far-field condition \eqref{far}, it follows that
\begin{equation} \label{mu -1 scaling}
\lambda \sim \mu^{-1} \tilde\lambda\quad \text{as} \quad \mu\to0.
\end{equation}
In light of the above uniform-flow approximation, the shear-continuity condition (\ref{interfacial}\textit{b}) is trivially satisfied at $\ord(\mu^{-1})$. 

\begin{figure}[t!]
\begin{center}
\includegraphics[scale=0.7]{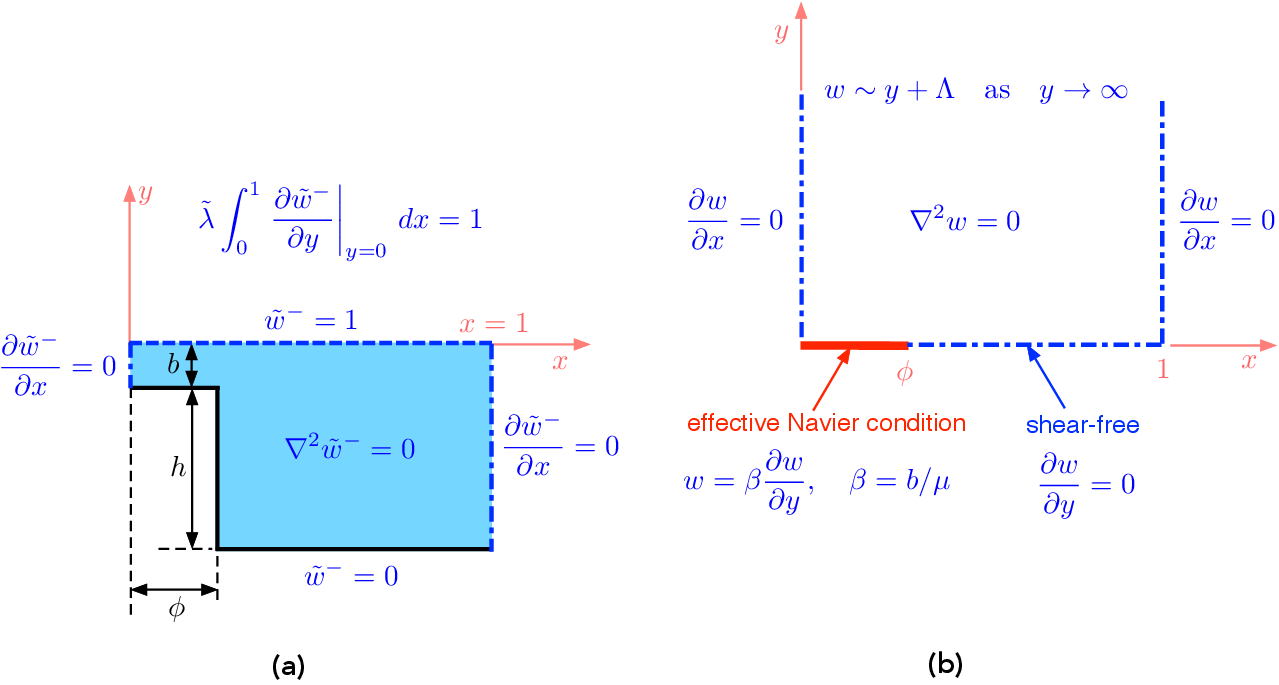}
\caption{(a) Interior problem derived in the limit $\mu\to0$ with $b$ fixed. (b) Generalized Philip problem derived in the limit $\mu\to0$ with $\beta=b/\mu$ fixed.}
\label{fig:key}
\end{center}
\end{figure}

In the interior region, we write
\begin{equation} \label{def tilde w}
w^- \sim \mu^{-1} \tilde\lambda\tilde w^-\quad \text{as} \quad \mu\to0,
\end{equation}
where we have factored out $\tilde\lambda$. The leading-order field $\tilde w^-$  
is governed by Laplace's equation and satisfies (i) the Dirichlet condition
\begin{equation} \label{Dirichlet tilde}
\tilde w^- = 1 \quad \text{at} \quad y=0,
\end{equation}
which follows from (\ref{interfacial}\textit{a}) and \eqref{w plus is tilde lambda}; 
(ii) the homogeneous Neumann condition [cf.~\eqref{sym per}]
\begin{equation}\label{sym tilde}
\pd{\tilde w^-}{x}=0 \quad \text{at} \quad x=0,1;
\end{equation}
and (iii) the no-slip condition at the solid boundary [cf.~\eqref{no slip}],
\begin{equation}\label{no slip tilde}
\tilde w^-=0,
\end{equation}
which introduces a dependence upon $b$, $\phi$ and $h$.
It is evident that $\tilde w^-$ 
 is fully determined by the above problem. Once evaluated, it is possible to calculate the associated ``drag'' (per unit length) acting on the lubricant in the longitudinal direction,
\begin{equation} \label{def cal F}
\mathcal D(b,\phi,h) = \int_0^1 \left.\pd{\tilde w^-}{y}\right|_{y=0} \, \dd x.
\end{equation}
The slip length then follows from \eqref{int in} in conjunction with \eqref{def tilde w},
\begin{equation} \label{slip is 1/D}
\tilde\lambda = \frac{1}{\mathcal D}.
\end{equation}
The interior problem governing $\tilde w^-$ and $\tilde\lambda$ is portrayed in Fig.~\ref{fig:key}(a).

We can now consider various limits of this interior problem.
One limit is that of thin ridges, $\phi\ll1$. It is evident that the problem governing $\tilde w^-$ is well-posed for $\phi=0$, with the domain simplifying to a rectangle, see Fig.~\ref{fig:key}(a). This regularity implies that we may simply substitute $\phi=0$ at leading order, 
\begin{equation} \label{slip is 1/D phi=0}
\tilde\lambda \sim  \frac{1}{\mathcal D(b,0,h)}\quad \text{as} \quad \phi\to0.
\end{equation}
Though we shall not pursue this calculation here, we note that $\mathcal D(b,0,h)$ may be obtained by solving the interior problem in this case using a Schwarz--Christoffel map \cite{Brown:book}. 

Another limit of interest is that of slight encapsulation, where $b$ is small. Given the conflict between the Dirichlet conditions \eqref{Dirichlet tilde} and \eqref{no slip tilde}, the problem governing $\tilde w^-$ is \emph{not} well-posed for $b=0$. To address the limit $b\ll1$ we conceptually decompose the interior domain into a ``thin-film'' region about the top of the ridge, where $0<x<\phi$ and $-b<y<0$, and the remaining ``cavity'' region. In the former, the velocity $\tilde w^-$ varies essentially in the $y$-direction, from $0$ at $y=-b$ to $1$ at $y=0$. The associated shear, $1/b$, is asymptotically larger than the $\ord(1)$ velocity gradient in the cavity. It is then evident that $\mathcal D$ is dominated by the lubrication contribution, namely $\mathcal D \sim \phi/b$. From \eqref{slip is 1/D} we therefore obtain
\begin{equation}
\tilde\lambda \sim \frac{b}{\phi} \quad \text{as} \quad b\to0. \label{small b}
\end{equation}

\section{Distinguished regimes in the  $\mu\ll1$ parameter space}
\label{sec:scalings}
Combining \eqref{mu -1 scaling} and \eqref{small b} gives (henceforth referred to as the algebraic approximation)
\begin{equation}\label{algebraic regime primitive} 
\lambda \sim \frac{b}{\mu\phi}.
\end{equation}
Note that \eqref{algebraic regime primitive} has been obtained by forming the limit $\mu\to0$ followed by the limit $b\to0$. The order of limits is crucial! Indeed, interchanging the order would lead to Philip's effective-slip problem discussed in the Introduction (corresponding to a grooved superhydrophobic surface); in that case, the slip length is a function of $\phi$ alone, unlike in \eqref{algebraic regime primitive}. The resolution to this discrepancy is that \eqref{algebraic regime primitive} breaks down when $b$ becomes comparable to $\mu$, since then the slip length (and thereby the fluid velocity) is no longer asymptotically large --- in contradiction to the premise underlying the interior problem. 

\begin{figure}[t!]
\begin{center}
\includegraphics[scale=0.6]{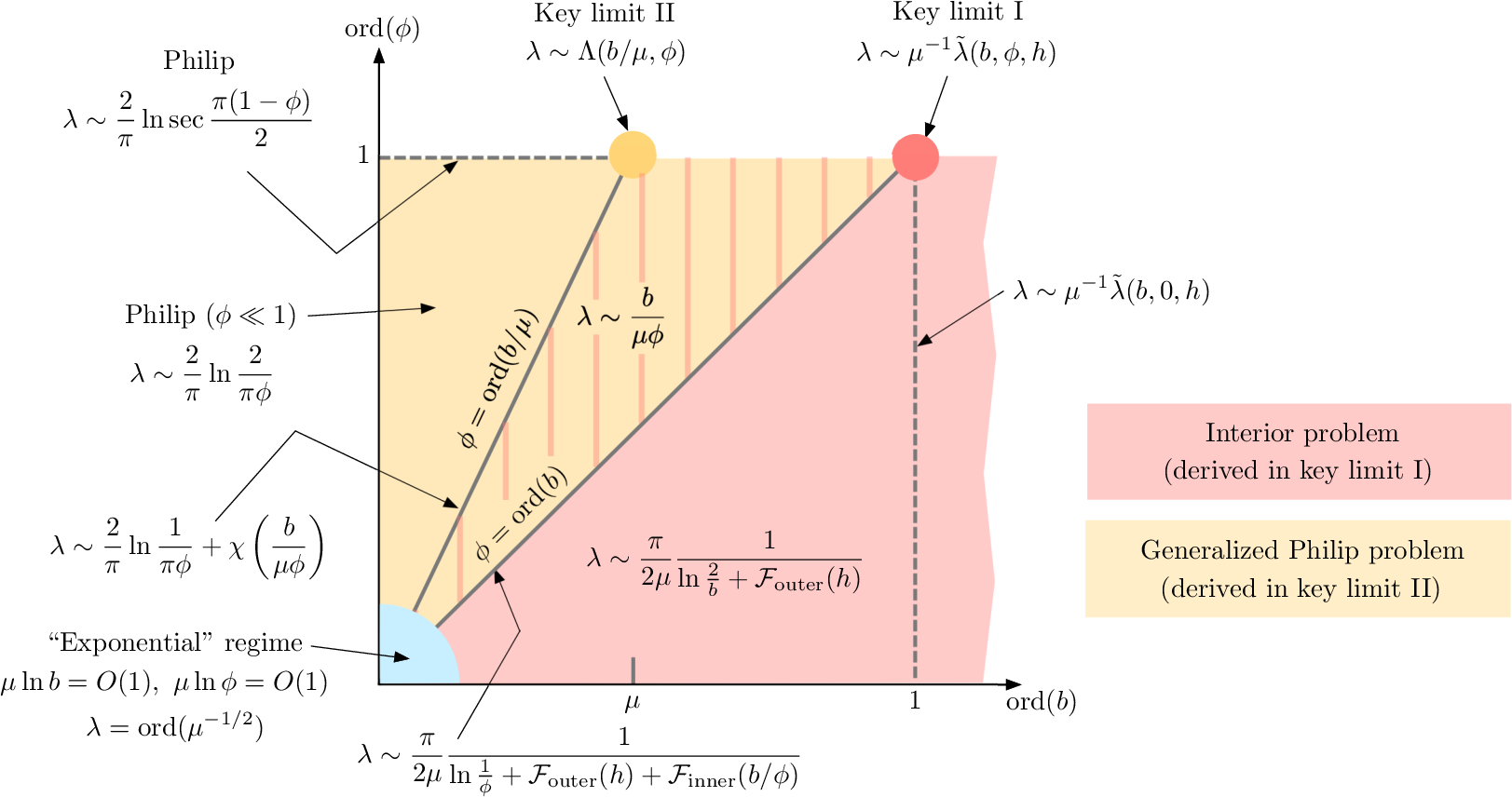}
\caption{``Phase map'' of the different regimes in the $(b,\phi)$ parameter space (for $\mu\ll1$).} 
\label{fig:Phase}
\end{center}
\end{figure}

We therefore identify two key limits as $\mu\to0$. 
The first, referred to hereafter as \textit{key limit} I, is where $b$ is held fixed; that limit has been addressed in the preceding section, leading to the interior problem described in Fig.~\ref{fig:key}(a). The second, referred to hereafter as \textit{key limit} II, is
where $b$ is comparable to $\mu$; this distinguished thin-film limit is the
subject of the next section. 

Rather than relying on the algebraic approximation \eqref{algebraic regime primitive}, the need for key limit II can be deduced more directly from a scaling argument employing the integral relation \eqref{int in}. Indeed, consider the drag integral on the left-hand side of that relation. For $b\ll1$,  the $\text{ord}(\mu \lambda/b)$ contribution to this integral from the thin-film region dominates the $\text{ord}(\mu \lambda)$ contribution from the cavity region; given \eqref{far}, these estimates assume $\lambda$ as the velocity scaling. It then follows from \eqref{int in} that $\lambda=\text{ord}(b/\mu)$, which must be large for consistency of the interior problem. Hence, as $b$ becomes comparable to $\mu$, with $\lambda$ approaching order unity, a new asymptotic regime must emerge. Physically, the amplified lubrication stresses in the thin-film region are then able to support the imposed shear without the need for large slip. 

Prior to addressing key limit II in the next section, we note that if \emph{both} $b$ and $\phi$ are small, the condition that \eqref{algebraic regime primitive} remains large, and so also the exterior velocity, is relaxed to $\mu\phi \ll b$. Hence, the domain of key limit I --- i.e., the domain in which the interior problem obtained in that limit yields a valid leading-order approximation as $\mu\to0$ --- expands from $\mu\ll b$ to $\phi\mu \ll b$. Within this expanded domain, the algebraic approximation \eqref{algebraic regime primitive} is expected to hold for $\phi\mu\ll b \ll \phi$. The upper asymptotic bound has to do with the lubrication geometry underlying that approximation; indeed, the lubricant film above the ridge tops ceases to be thin as $\phi$ becomes comparable to $b$. This suggests a new distinguished limit, within the expanded domain of key limit I, where $b/\phi$ is held fixed. The analysis of the interior problem in that distinguished sub-limit is postponed to Sec.~\ref{sec:thin}. 

Fig.~\ref{fig:Phase} shows an asymptotic ``phase map'' of the the different regimes in the $(b,\phi)$ parameter space (for $\mu\ll1$). 
Among the various scalings and approximations depicted, we have already identified key limits I and II, which are marked by solid circles; the domain of key limit I; the domain of the algebraic approximation \eqref{algebraic regime primitive}; and the regular small-$\phi$ approximation \eqref{slip is 1/D phi=0} of the interior problem. The remaining results depicted in the phase map will be developed in Secs.~\ref{sec:b is mu} and \ref{sec:thin}; we shall return to discuss this map in Sec.~\ref{Sec:Discussion}. 

\section{Generalized Philip problem for thin films} 
\label{sec:b is mu}

In this section we address the distinguished thin-film limit 
\begin{equation}
\label{II}
\mu,b\to0 \quad \text{with}\quad \text{$b/\mu$ fixed},
\end{equation}
which corresponds to key limit II. 
To that end, we define 
\begin{equation}
\beta = \frac{b}{\mu} 
\label{def beta}
\end{equation}
and rewrite \eqref{II} as 
\begin{equation}
\label{dist II}
\mu\to0 \quad \text{with} \quad\text{$\beta$ fixed}.
\end{equation}
The scaling arguments given in Sec.~\ref{sec:scalings} suggest that the velocity fields $w^{\pm}$ and the slip length $\lambda$ are $\text{ord}(1)$ in this limit. We accordingly posit the approximations
\begin{equation} \label{Lambda}
w^{\pm} \sim w_0^{\pm}, \quad \lambda \sim \Lambda \quad \text{as} \quad \mu\to0 \quad \text{with $\beta$ fixed}.
\end{equation}
In what follows, we analyze the flow problem in the limit \eqref{dist II}, seeking to derive a reduced problem governing the leading-order slip length $\Lambda$. 
\subsection{Exterior approximation}
We start with the exterior domain. The leading-order velocity field $w_0^+$ satisfies Laplace's equation; the symmetry-periodicity conditions 
\begin{equation}
\label{u sym per}
\pd{w_0^+}{x} = 0 \quad \text{at} \quad x=0,1;
\end{equation}
and the far-field condition [cf.~\eqref{far}]
\begin{equation}\label{far in Lambda}
w_0^+ = y + \Lambda + o(1) \quad \text{as} \quad y\to\infty.
\end{equation}

The condition satisfied by $w_0^+$ at $y=0$ remains to be specified. To that end, we  consider the flow of the lubricant. Since the length scale of the cavity region is of order unity, consideration of the interfacial shear balance (\ref{interfacial}\textit{b}) in the limit $\mu\ll1$ implies the effective shear-free boundary condition
\begin{equation}
\label{u shear free}
\pd{w_0^+}{y}=0 \quad \text{at} \quad y=0, \quad \phi<x<1.
\end{equation}
To derive the effective condition for $0< x< \phi$, we turn to the thin-film region. 

\subsection{Thin-film region}
The thin-film region is studied by holding $x$ and the strained coordinate $Y=y/\mu$ fixed. It is then given by $-\beta\le Y\le 0$ and $0\le x<\phi$. In that region we define $W^-(x,Y;\mu)=w^-(x,y;\mu)$, and posit the approximation 
\begin{equation}
W^-\sim W_0^- \quad \text{as} \quad \mu\to0 \quad \text{with $\beta$ fixed}.
\end{equation}
The leading-order velocity $W_0^-$  is governed by the differential equation
\begin{equation}\label{film eq lo}
\pd{^2W_0^-}{Y^2}=0,
\end{equation}
which follows from a leading-order balance of Laplace's equation. In addition it satisfies (i)
the no-slip condition 
\begin{equation}\label{film no slip}
W_0^-=0 \quad \text{at} \quad Y=-\beta;
\end{equation}
(ii) the symmetry condition
\begin{equation}\label{film symmetry}
\pd{W_0^-}{x}=0 \quad \text{at} \quad x=0;
\end{equation}
and (iii) the interfacial conditions
\refstepcounter{equation}
$$
\label{film interfacial}
\left.W_0^-\right|_{Y=0}=\left.w_0^+\right|_{y=0}, \quad \left.\pd{W_0^-}{Y}\right|_{Y=0}=\left.\pd{w_0^+}{y}\right|_{y=0},
\eqno{(\theequation{\mathit{a},\mathit{b}})}
$$
which follow from \eqref{interfacial}.

Integrating \eqref{film eq lo} gives, upon using \eqref{film no slip}, 
\begin{equation}
\label{Wm0 sol}
 W_0^-(x,Y)=A(x)(Y+\beta).
\end{equation}
Substituting \eqref{Wm0 sol} into the stress balance (\ref{film interfacial}b), we find
\begin{equation}
\label{A sol}
A(x) = \left.\pd{ w_0^+}{y}\right|_{y=0}. 
\end{equation}
(The symmetry condition \eqref{u sym per} satisfied by $ w_0^+$ at $x=0$ ensures that the derivative of $A(x)$ vanishes there, so that the leading-order balance of the symmetry condition \eqref{film symmetry} is also satisfied.)
With \eqref{A sol}, the continuity condition (\ref{film interfacial}a) yields the Robin-type condition
\begin{equation}\label{NS}
 w_0^+ = \beta \pd{ w_0^+}{y} \quad \text{at} \quad y=0, \quad 0< x <\phi. 
\end{equation}

\subsection{Generalized Philip problem}
\label{ssec:generalized}
We have reduced the original problem to an exterior problem in the domain $y>0$. We find it convenient to recapitulate the latter problem, with the leading-order exterior velocity $w_0^+$ now simply relabeled as $w$. The notation $\Lambda$ for the leading-order slip length is retained in order to distinguish this approximation from the exact slip length $\lambda$.  

The problem governing $w$ consists of: (i) Laplace's equation,
\begin{equation} \label{Laplace II}
\pd{^2w}{x^2}+\pd{^2w}{y^2} = 0;
\end{equation}
(ii) the symmetry--periodicity conditions 
\begin{equation}
\label{tilde sym per}
\pd{ w}{x} = 0 \quad \text{at} \quad x=0,1;
\end{equation}
(iii) the far-field condition 
\begin{equation}
\label{u far}
 w(x,y) = y + {\Lambda} +o(1) \quad \text{as} \quad y\to\infty;
\end{equation}
(iv) the effective shear-free boundary condition [cf.~\eqref{u shear free}]
\begin{equation}
\label{tilde shear free}
\pd{ w}{y}=0 \quad \text{at} \quad y=0, \quad \phi<x<1;
\end{equation}
and (v) the Navier condition [cf.~\eqref{NS}]
\begin{equation}\label{tilde NS}
 w = \beta \pd{ w}{y} \quad \text{at} \quad y=0, \quad 0< x <\phi. 
\end{equation}

The resulting problem is depicted in Fig.~\ref{fig:key}(b). It is independent of $h$, and involves two parameters: $\phi$, which may be interpreted as the apparent solid fraction, and $\beta$. It is easy to see that it implies the integral relation 
\begin{equation}\label{MP int1}
\int_0^\phi\left.\pd{ w}{y}\right|_{y=0}\,\dd x = 1,
\end{equation} 
which can also be deduced from a leading-order balance of the exact integral relation \eqref{int out}. Using \eqref{tilde NS} we obtain the equivalent constraint, 
\begin{equation}\label{MP int2}
\int_0^\phi  \left.w\right|_{y=0} \,\dd x = \beta. 
\end{equation}

For $\beta=0$, the Navier condition \eqref{tilde NS} simplifies to a no-slip condition, whereby we recover Philip's effective-slip problem corresponding to a grooved superhydrophobic surface. We accordingly refer to the problem \eqref{Laplace II}--\eqref{tilde NS} as a \emph{generalized Philip problem}. From Philip's solution, we have that \cite{Philip:72,Philip:72:integral}
\begin{equation}
\Lambda(\beta,\phi) \sim \frac{2}{\pi}\ln\sec\frac{\pi(1-\phi)}{2} \quad \text{as} \quad \beta\to0.
\label{Philip}
\end{equation}

\begin{figure}[b]
\begin{center}
\includegraphics[scale=0.5]{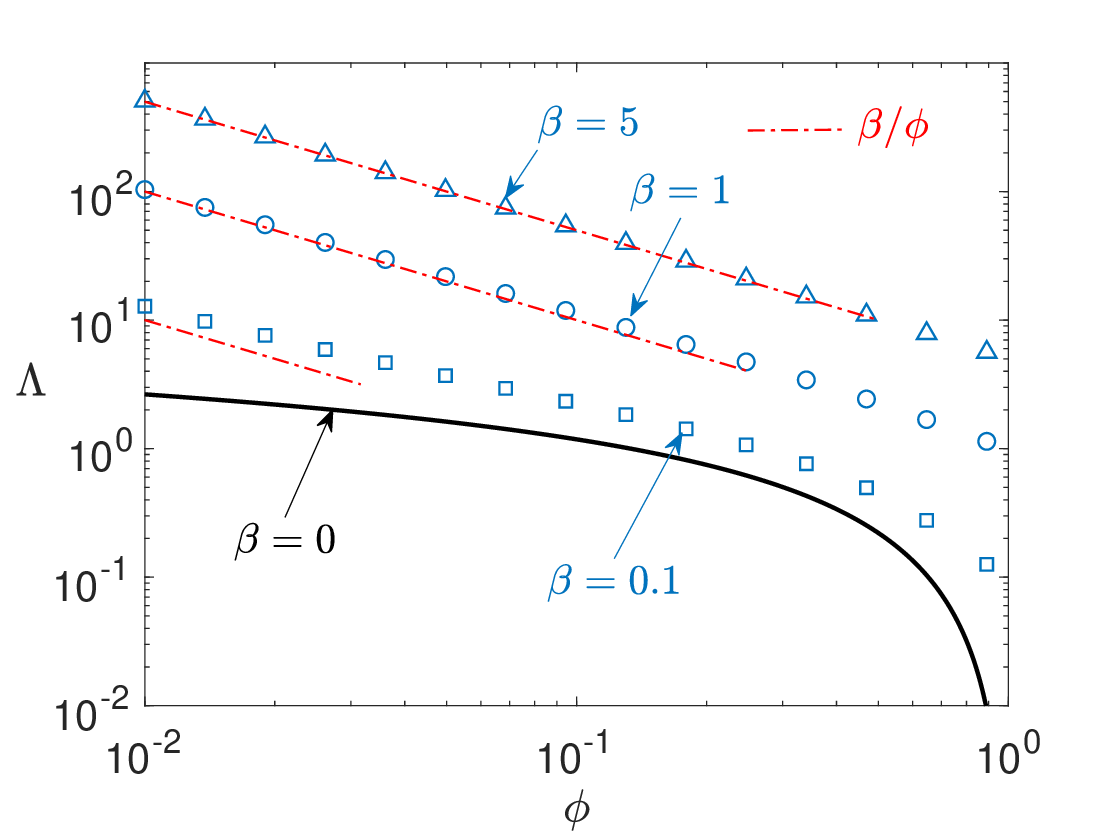}
\caption{Solutions of the generalized Philip problem, showing the leading-order slip length $\Lambda$ as a function of the ridge semi-width (or apparent solid fraction) $\phi$, for several values of $\beta=b/\phi$. Symbols: numerical solutions. Solid curve: Philip's result \eqref{Philip} for $\beta=0$. Dash-dotted curves: the algebraic approximation $\Lambda\sim \beta/\phi$, valid for $\beta\gg\phi$ (see subsection \ref{ssec:algebraicPhilip}).} 
\label{fig:modNav}
\end{center}
\end{figure}
\subsection{Numerical solutions}
\label{ssec:numericalofNavier}
The Navier condition is not invariant under a conformal mapping, so Philip's solution approach \citep{Philip:72} does not apply. It is, however, trivial to solve the generalized Philip problem numerically. We seek a solution in the form of the Fourier series 
\begin{equation}
 w(x,y) = y + \Lambda + \sum_{n=1}^{\infty}a_ne^{-\pi n y} \cos(\pi n x),
\end{equation}
where $\{a_n\}_{n=1}^\infty$ are real-valued constants. This general solution satisfies Laplace's equation, the symmetry-periodicity conditions \eqref{tilde sym per} and the far-field condition \eqref{u far}. The coefficients $\{a_n\}_{n=1}^\infty$ and slip length $\Lambda$ are set by the mixed boundary conditions \eqref{tilde shear free}--\eqref{tilde NS}, 
which here become
\begin{subequations}
\begin{gather}
\Lambda-\beta + \sum_{n=1}^\infty a_n(1+\pi n \beta)\cos (\pi n x)=0 \quad \text{for} \quad 0< x<\phi,\\
1-\pi\sum_{n=1}^{\infty}na_n\cos(\pi n x)=0 \quad \text{for}\quad\phi<x< 1. 
\end{gather}
\end{subequations}
We satisfy these conditions by numerical collocation. 
The resulting slip length $\Lambda$ is shown in Fig.~\ref{fig:modNav} as a function of $\phi$ for the indicated values of $\beta$.  

\subsection{Algebraic regime}
\label{ssec:algebraicPhilip}
The numerical results in Fig.~\ref{fig:modNav} show the slip length $\Lambda$ increasing with $\beta$, starting from Philip's limit \eqref{Philip}. This motivates a large-$\beta$ analysis of the generalized Philip problem. In that limit, the  Navier condition appears to reduce to a shear-free condition; the entire boundary $y=0$ is then shear-free. This singular scenario suggests that $w$ is asymptotically large and, to leading order, uniform; given the far-field condition \eqref{u far}, this uniform velocity coincides with the slip length $\Lambda$. The integral relation in the form \eqref{MP int2} thus  implies the approximation 
\begin{equation}
\label{philip large beta}
{\Lambda} \sim \frac{\beta}{\phi} \quad \text{for} \quad \beta\gg1.
\end{equation}
Unsurprisingly, it coincides with \eqref{small b}, the small-$b$ limit of the interior problem --- this is the algebraic approximation discussed in Sec.~\ref{sec:scalings}. Thus, with increasing $\beta$, the generalized Philip problem describes the transition from Philip's limit \eqref{Philip} corresponding to an inviscid lubricant without encapsulation, to the large $\text{ord}(\mu^{-1})$ slip lengths for nearly inviscid lubricants and moderate encapsulation heights. The approximation \eqref{philip large beta} is depicted in Fig.~\ref{fig:modNav} in order to illustrate this transition.

In Sec.~\ref{sec:scalings}, we showed that with $\phi$ varying the algebraic approximation holds for $\mu\phi \ll b\ll\phi$, where the lower bound arised from the collapse of the interior problem and the upper bound from a requirement that the lubricant film wetting the ridge is thin. 
In terms of $\beta$, this range becomes $\phi\ll\beta \ll \phi/\mu$. From the diametric viewpoint of key limit II, the upper bound comes from the collapse of the generalized Philip problem, since the derivation of that problem hinges upon the thin-film requirement. The lower bound can be deduced from an analysis of the generalized Philip problem. Indeed, $\phi\ll1$ suggests that the boundary $y=0$ is again shear-free. Similarly to above, this singular scenario suggests an asymptotically large uniform flow, whereby the integral relation \eqref{MP int2} leads to the algebraic approximation $\beta/\phi$; the bound follows by demanding consistency, namely that this ratio is large \footnote{This argument may appear naive, as the integral relation is now applied to an asymptotically small interval. Nonetheless, the uniform-flow approximation remains consistent; indeed, the Navier condition \eqref{tilde NS} suggests that an $\mathrm{ord}(\beta/\phi)$ large uniform velocity in the $\mathrm{ord}(\phi)$ vicinity of that interval generates only weaker $\mathrm{ord}(1)$ velocity variations.}.

Whereas the algebraic approximation $\beta/\phi$ exhibits an algebraic (power-law) singularity as $\phi\to0$, Philip's result \eqref{Philip} scales logarithmically in that limit. The transition with increasing $\beta$  from  logarithmic to algebraic growth as $\phi\to0$ is apparent in Fig.~\ref{fig:modNav}. We expect this transition to occur for $\beta=\text{ord}(\phi)$, when the algebraic approximation breaks down. This distinguished sub-limit of the generalized Philip problem is considered next.

\subsection{Transition from logarithmic to algebraic small-solid-fraction singularity}
\label{subsec:Transition}
To address the case $\beta=\text{ord}(\phi)$, we define,
\begin{equation}
\kappa=\frac{\beta}{\phi}. \label{def kappa}
\end{equation}
Starting from the generalized Philip problem \eqref{Laplace II}--\eqref{tilde NS}, we consider the limit $\phi\to0$ with $\kappa$ fixed. While the interval associated with the Navier condition \eqref{tilde NS} shrinks to a point in that limit, condition \eqref{MP int1} necessitates a finite force from that interval. This suggests the need for matched asymptotic expansions, where said interval is represented by a point singularity in the ``outer'' region, while the ``inner'' region is on the scale of that interval. The point singularity demanded by \eqref{MP int1} is a line force parallel to the grooves; in terms of the harmonic potential $w$, this corresponds to a two-dimensional source singularity. Accordingly, we anticipate velocities scaling with the logarithm of $\phi$.

We follow the modern convention in matched asymptotic expansions \citep{Fraenkel:69}, where terms of similar algebraic order are grouped together. Thus, we posit the expansion 
\begin{equation} \label{def bar Lambda}
{\Lambda}(\phi) \sim \bar{\Lambda}(\phi) \quad \text{as} \quad \phi\to 0, 
\end{equation}
wherein $\bar{\Lambda}$ is permitted to vary logarithmically with $\phi$. The asymptotic correction is accordingly rendered ``algebraically small'' --- asymptotically smaller than some positive power of $\phi$.

\subsubsection{Outer region}
Similarly to \eqref{def bar Lambda}, we posit the outer approximation
\begin{equation}
 w(x,y;\phi) \sim \bar w(x,y; \phi) \quad \text{as} \quad \phi\to 0,
\end{equation}
where the leading-order velocity $\bar w$ is allowed to vary logarithmically with $\phi$. It satisfies Laplace's equation in the semi-strip $0<x<1$ with $y>0$; the far-field condition in the form 
\begin{equation}
\bar w(x,y) \sim y + \bar{\Lambda}  \quad \text{as} \quad y\to\infty; 
\end{equation}
and the Neumann condition,
\begin{equation}\label{bar w n}
\pd{\bar w}{n} = 0,
\end{equation}
on the semi-strip boundaries $x=0,1$ [cf.~\eqref{tilde sym per}] and $y=0$  [cf.~\eqref{tilde shear free}]. 

The Neumann condition \eqref{bar w n} does not apply at the origin. In principle, the appropriate behavior near the origin follows from the need to satisfy asymptotic matching with the inner region. It proves sufficient, however, to exclude any algebraic growth as $r\to0$, where $r=\sqrt{x^2+y^2}$; indeed, such a growth would render an algebraically large velocity in the inner region. The integral relation \eqref{MP int1} then implies the behavior
\begin{equation} \label{source}
\bar w(x,y) \sim \frac{2}{\pi}\ln r  \quad \text{as} \quad r \to 0,
\end{equation}
corresponding to the potential source singularity anticipated above. 

The solution to the resulting problem for $\bar{w}$ is well known \citep{Yariv:23,Peng:25}; it provides the following refinement of the origin singularity \eqref{source}, 
\begin{equation}\label{u0 far extended}
\bar w(x,y) \sim \frac{2}{\pi}\ln r +\frac{2}{\pi}\ln\pi + \bar{\Lambda} \quad \text{as} \quad r\to 0,
\end{equation}
where the asymptotic error is algebraically small.

At this stage, $\bar{\Lambda}$ is still unknown. We now turn to the inner region on the scale $\phi$ about the origin. 
\subsubsection{Inner region}
To study the inner region we introduce the stretched coordinates 
\refstepcounter{equation}
$$
\hat x=x/\phi, \quad \hat y=y/\phi,
\eqno{(\theequation{\mathit{a},\mathit{b}})}
$$
and consider the inner limit where $\phi\to0$ with $\hat x,\hat y$ fixed. 
We expand 
\begin{equation}
 w(x,y;\phi) \sim \hat w(\hat x,\hat y;\phi) \quad \text{as} \quad \phi\to0,
\end{equation}
allowing for a logarithmic dependence of $\hat w$ upon $\phi$. 
The leading-order inner velocity is governed by Laplace's equation in the first quadrant. In addition, it satisfies the symmetry condition
\begin{equation} \label{symmetry hat}
\pd{\hat w}{\hat x}=0 \quad \text{at} \quad \hat x=0,
\end{equation}
as well as the mixed boundary conditions at $\hat y=0$:
\begin{subequations}
\begin{gather}
\hat w=\kappa \pd{\hat w}{\hat y}\quad \text{for}  \quad 0< \hat x<1, 
\label{Robin} \\
\pd{\hat w}{\hat y}=0\quad \text{for}  \quad  \hat x>1. 
\label{shear free hat}
\end{gather}
\end{subequations}

We note that the integral balance \eqref{MP int1} becomes
\begin{equation}\label{U0 int}
\int_0^1\left.\pd{\hat w}{\hat y}\right|_{\hat y=0}\,\dd\hat x=1,
\end{equation}
or, using \eqref{Robin}, 
\begin{equation}\label{U0 int other}
\int_0^1\left.\hat w\right|_{\hat y=0}\,\dd\hat x=\kappa.
\end{equation}
The flux \eqref{U0 int} implies the far-field condition
\begin{equation}\label{U far}
\hat w(\hat x,\hat y) \sim \frac{2}{\pi}\ln \hat r \quad \text{as} \quad \hat r\to\infty,
\end{equation}
where $\hat r=\sqrt{\hat x^2+\hat y^2}$. [Alternatively, \eqref{U far} follows from asymptotic matching with \eqref{source}.]

Condition \eqref{U far} closes the problem governing $\hat w$, which depends upon the single parameter $\kappa$. Out interest is in the uniform correction to the far-field \eqref{U far}, i.e.,
\begin{equation} \label{def chi}
\chi(\kappa)=\lim_{\hat r\to\infty}\left(\hat w-\frac{2}{\pi}\ln \hat r\right),
\end{equation}
which is an output of the $\hat w$ problem. Given definition \eqref{def chi} we may modify \eqref{U far} to
\begin{equation}\label{U far modified}
\hat w(\hat x,\hat y) = \frac{2}{\pi}\ln \hat r + \chi(\kappa) + o(1) \quad \text{as} \quad \hat r\to\infty.
\end{equation}
Asymptotic matching with \eqref{u0 far extended} provides the requisite slip length $\bar\Lambda$ in terms of the offset $\chi$,
\begin{equation}
\label{slip in Lambda}
\bar{\Lambda} = \frac{2}{\pi}\ln \frac{1}{\pi\phi} + \chi(\kappa),
\end{equation}
thereby separating the dependence upon $\phi$ and $\kappa$. 

To solve the $\hat w$-problem, it is convenient to employ elliptical coordinates $(u,v)$, defined \citep{Moon:book} by the map $\hat x+\ii\hat y=\cosh(u+\ii v)$. With 
\begin{equation}
\frac{{\hat x}^2}{\cosh^2 {u}}+\frac{{\hat y}^2}{\sinh^2 {u}} = 1,
\end{equation}
we see that ${u}$ vanishes for $0< \hat x<1$ and $\hat y=0$, and behaves as $\ln \hat r+\ln 2+o(1)$ as $\hat r\to\infty$. 
Let $\varphi({u},{v})=\hat w(\hat x,\hat  y)$. It is governed by Laplace equation in a semi-strip,
\begin{equation}
\pd{^2\varphi}{u^2} + \pd{^2\varphi}{v^2} =0 \quad \text{for} \quad 0<u<\infty, \quad 0<v<\frac{\pi}{2}, \label{Laplace elliptic}
\end{equation}
and satisfies: (i) the Neumann conditions
\refstepcounter{equation}
$$
\label{Neumann phi}
\pd{\phi}{v}=0\quad \text{at} \quad v=0, \qquad
\pd{\phi}{v}=0\quad \text{at} \quad v=\frac{\pi}{2},
\eqno{(\theequation{\mathit{a},\mathit{b}})}
$$
which follows from \eqref{symmetry hat} and \eqref{shear free hat}, respectively;
(ii) the Navier-slip condition
\begin{equation}
\label{Navier elliptical}
\varphi  = \frac{\kappa}{\sin v} \pd{\varphi}{{u}} \quad \text{at} \quad {u}=0,
\end{equation}
which follows from \eqref{Robin};
and (iii) the far-field condtion,
\begin{equation}
\varphi = \frac{2}{\pi}{u}+\chi(\kappa)-\frac{2}{\pi}\ln 2 + o(1) \quad \text{as} \quad {u}\to\infty,
\label{varphi far elliptical}
\end{equation}
which follows from \eqref{U far modified}.

\subsubsection{Asymptotic limits and collocation solution}
For large $\kappa$, \eqref{Robin} and \eqref{U0 int other} imply the uniform flow 
\begin{equation}
\hat w \sim \kappa.
\end{equation}
It follows from \eqref{U far modified} that
\begin{equation} \label{chi large kappa}
\chi(\kappa)\sim \kappa \quad \text{as} \quad \kappa\to\infty,
\end{equation}
 whereby \eqref{slip in Lambda} yields
\begin{equation}
\bar{\Lambda} \sim \kappa +  \frac{2}{\pi}\ln \frac{1}{\phi}\quad \text{as} \quad \kappa\to\infty,
\end{equation}
where the relative magnitude of the two terms is indeterminate. The leading term coincides with the algebraic behavior already encountered in \eqref{philip large beta}.
\begin{figure}[hbtp]
\begin{center}
\includegraphics[scale=0.5]{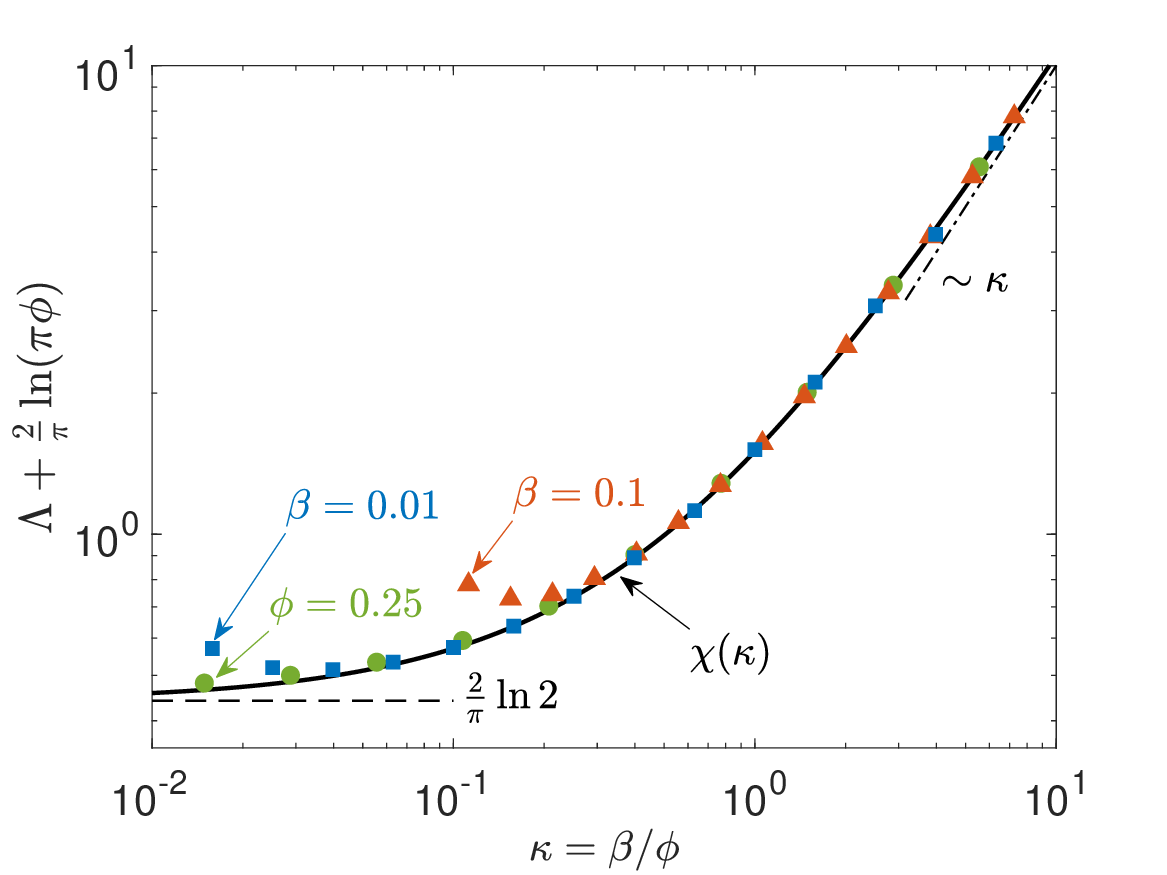}
\caption{Collapse of numerical solutions of the generalized Philip problem on the universal function $\chi(\kappa)$ obtained in the distinguished sub-limit of that problem, $\phi\to0$ with $\kappa=\beta/\phi$ held fixed. The numerical solutions were carried out for both fixed $\phi$ (circles) and fixed $\beta$ (squares and triangles).}
\label{fig:MPcollapse}
\end{center}
\end{figure}

For small $\kappa$, condition \eqref{Navier elliptical} reduces to a homogeneous Dirichlet condition. From \eqref{Laplace elliptic}--\eqref{varphi far elliptical}, we readily obtain
\begin{equation}
\varphi \sim \frac{2}{\pi}{u}. 
\end{equation}
It also follows from \eqref{varphi far elliptical} that
\begin{equation}  \label{chi small kappa}
\chi(\kappa)\sim \frac{2}{\pi}\ln 2 \quad \text{as} \quad \kappa\to0. 
\end{equation}
The associated slip length follows from \eqref{slip in Lambda} as
\begin{equation}
\bar{\Lambda} \sim \frac{2}{\pi}\ln \frac{2}{{\pi} \phi} \quad \text{as} \quad \kappa\to0.
\label{Philip log}
\end{equation}
This corresponds to the familiar small-solid-fraction approximation of Philip's formula \eqref{Philip} for superhydrophobic surfaces. 

Consider now arbitrary $\kappa$. The general solution of \eqref{Laplace elliptic} that satisfies \eqref{Neumann phi} and \eqref{varphi far elliptical} 
is obtained using Fourier series, 
\begin{equation} \label{Fourier series}
\varphi=\frac{2}{\pi}{u} + \chi(\kappa)-\frac{2}{\pi}\ln 2 + \sum_{n=1}^{\infty} t_n e^{-2n{u}}\cos(2n{v}),
\end{equation}
where $\{t_n\}_{n=1}^{\infty}$ are real-valued constants. The coefficients $\{t_n\}_{n=1}^{\infty}$ and $\chi$  are determined by imposing condition \eqref{Navier elliptical}.  
In principle, one can apply Fourier orthogonality to derive an infinite algebraic system governing these coefficients. Instead, we used collocation at ${v}_j=(\pi/2)j/N$, $j=0,1,2,\ldots,N$, to solve approximately for $\{t_n\}_{n=1}^{N}$ together with $\chi(\kappa)$. The resulting $\chi(\kappa)$ is plotted in Fig.~\ref{fig:MPcollapse}, together with the asymptotic approximations
\eqref{chi large kappa} and \eqref{chi small kappa}. We also plot $\Lambda+(2/\pi)\ln(\pi\phi)$ [cf.~\eqref{slip in Lambda}], where $\Lambda$ is determined from the numerical solution of the generalized Philip problem, carried out for both fixed $\phi$ and fixed $\beta$. It is evident that the numerical data collapses upon the universal function $\chi(\kappa)$.

\section{Slightly encapsulated thin ridges}\label{sec:thin}
We return to the interior problem derived in key limit I (see Sec.~\ref{sec:nearly}). This problem governs the rescaled leading-order interior velocity $\tilde{w}^-$ and slip length $\tilde{\lambda}$, which depend on the parameters $b$, $\phi$ and $h$. So far, the only explicit solution we have obtained to this problem is the algebraic approximation $\tilde{\lambda}\sim b/\phi$, valid for lubrication geometries satisfying $b\ll \phi$. (Recall from Sec.~\ref{sec:scalings}, however, that the interior problem and the algebraic approximation cease to hold when $b$ becomes comparably small to $\mu\phi$.) In this section, we analyze the distinguished sub-limit of the interior problem where $b$ and $\phi$ are comparably small. Defining [cf.~\eqref{def beta}]
\begin{equation}
\mathcal B = \frac{b}{\phi},
\label{def B}
\end{equation}
this limit becomes  
\begin{equation} \label{dist thin}
\phi\to0 
\quad \text{with } \mathcal B\text{ fixed}.
\end{equation}

The lubrication geometry underlying the algebraic approximation is lost: the small thickness of the film wetting the ridge top is now comparable to the small ridge width. This suggests that the domain of the interior problem asymptotically decomposes into an outer region, on the order-unity scale of the cavity --- where the finite width of the ridge is indiscernible --- and an inner region corresponding to the $\text{ord}(\phi)$ vicinity of the ridge top. Given the Dirichlet boundary condition \eqref{Dirichlet tilde}, we expect $\tilde{w}^{-}=\text{ord}(1)$ in both the inner and outer regions. The geometry then suggests that $\partial\tilde{w}^-/\partial y$ is $\text{ord}(\phi^{-1})$ in the inner region and $\text{ord}(1)$ in the outer region; thus, noting the $\text{ord}(\phi)$ extent of the inner region, we estimate that both regions contribute at order unity to the drag $\mathcal{D}$, defined by the integral in \eqref{def cal F}. This suggests that $\mathcal{D}$ is, in fact, logarithmically large in $\phi$ --- with the logarithmic enhancement associated with the overlap between the inner and outer regions \cite{Hinch:book}. Since $\tilde{\lambda}=1/\mathcal{D}$ [cf.~\eqref{def cal F}], we accordingly expect $\tilde{\lambda}$ to be logarithmically small in $\phi$. 

\subsection{Outer problem}
The outer region is analyzed by considering the distinguished sub-limit \eqref{dist thin}, holding $x$ and $y$ fixed within the rectangle $0<x<1$ and $-h<y<0$. Given our above estimate that $\tilde{w}^-=\text{ord}(1)$, we posit the leading-order outer approximation
\begin{equation}
\label{w- thin}
\tilde w^-(x,y;\phi) \sim w^*(x,y) \quad \text{as} \quad \phi\to0.
\end{equation}

The velocity $w^*$ satisfies: (i) Laplace's equation,
\begin{equation}\label{laplace w start}
\pd{^2w^*}{x^2} + \pd{^2w^*}{y^2}=0;
\end{equation}
(ii) the no-slip conditions [cf.~\eqref{no slip tilde}]
\begin{equation}\label{no slip bottom -1}
w^*=0 \quad \text{at} \quad y=-h \quad \text{for} \quad 0<x<1,
\end{equation}
and 
\begin{equation}\label{no slip -1}
w^*=0  \quad \text{at} \quad x=0 \quad \text{for} \quad -h<y<0;
\end{equation}
(iii) the periodicity condition [cf.~\eqref{sym tilde}]
\begin{equation}\label{sym per -1}
\pd{w^*}{x}=0 \quad \text{at} \quad x=1 \quad \text{for} \quad -h<y<0;
\end{equation}
and (iv) the Dirichlet condition [cf.~\eqref{Dirichlet tilde}]
\begin{equation}\label{Dirichlet -1}
w^* = 1 \quad \text{at} \quad y=0 \quad \text{for} \quad 0<x<1.
\end{equation}
\begin{figure}[t!]
\begin{center}
\includegraphics[scale=0.38]{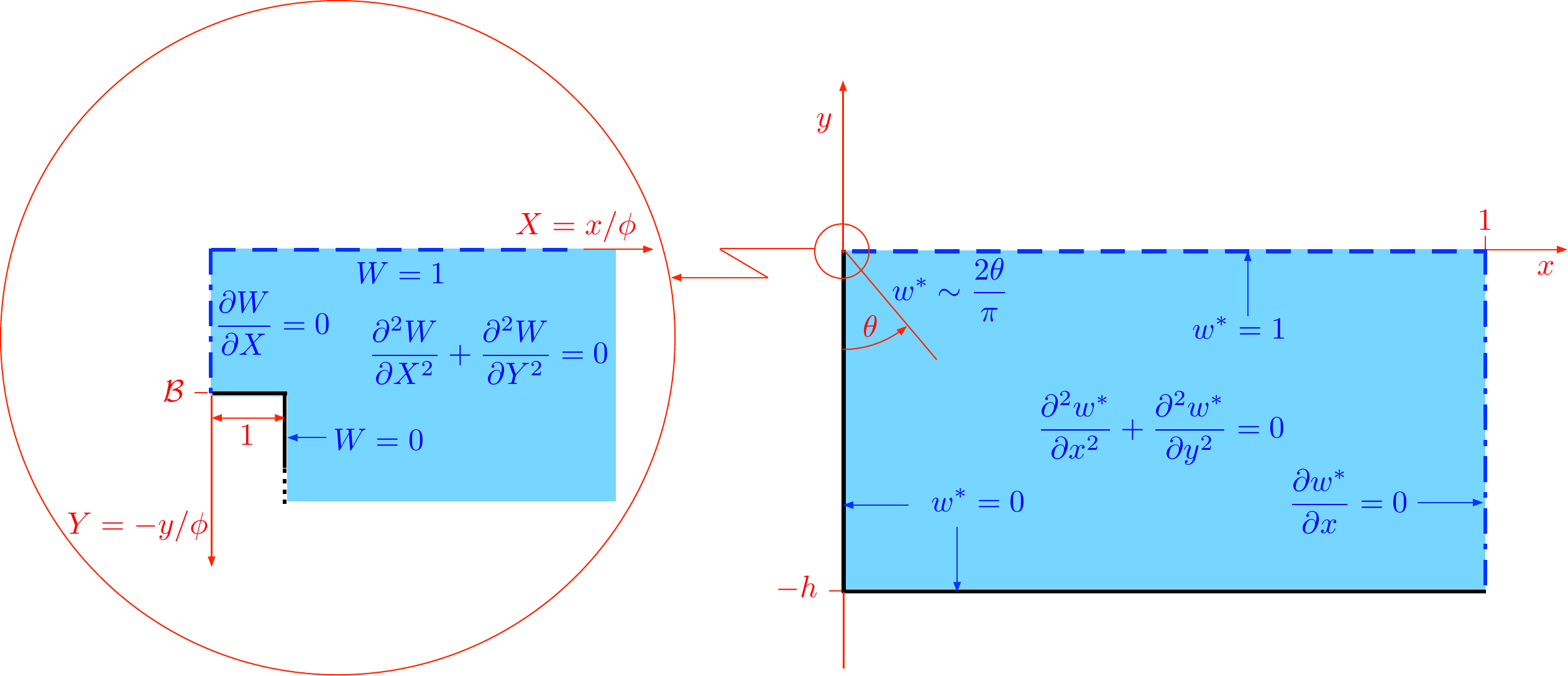}
\caption{The interior problem in the distinguished sub-limit where $\phi\to0$ with $\mathcal B=b/\phi$ fixed. The inset depicts the $\phi$-scale inner region near the origin.}
\label{fig:interior}
\end{center}
\end{figure}

In principle, determining the behavior of $w^*$ near the origin involves asymptotic matching with the inner region. It proves sufficient to require that $w^*$ is bounded near the origin, in accordance with our expectation that $\tilde{w}^-$ is also $\text{ord}(1)$ in the inner region. In that case, a local analysis near the origin, using the boundary conditions \eqref{no slip -1} and \eqref{Dirichlet -1}, yields
\begin{equation}
\label{local}
w^* \sim \frac{2\theta}{\pi} \quad \text{as} \quad r\to0, 
\end{equation}
where $\theta$ is the polar angle measured in the clockwise direction from the negative $y$-axis. This corner condition closes the outer problem governing $w^*$, which is depicted in Fig.~\ref{fig:interior}. We note that this problem  depends solely upon $h$. 

It may appear that we could obtain a leading-order approximation for $\mathcal D$ from \eqref{def cal F} by replacing $\partial\tilde w^-/\partial y$ by $\partial w^*/\partial y$ in the integral appearing therein. However, the corner behavior \eqref{local} is associated with the derivative behavior
\begin{equation}
\pd{w^*}{y}(x,0) \sim \frac{2}{\pi x} \quad \text{as} \quad x\searrow0,
\label{1/x}
\end{equation}
meaning that said integral would diverge logarithmically. This agrees with our expectation that both the inner and outer region contribute to $\mathcal{D}$ at leading order. 

The solution of the problem governing $w^*$ was carried out by Peng \textit{et al.} \cite{Peng:25} using a Schwarz--Christoffel mapping. The specific result we will need is the regularized drag integral 
\begin{equation}
\int_0^1\left[\pd{w^*}{y}(x,0)-\frac{2}{\pi x}\right]\,\dd x= \frac{2}{\pi}\mathcal F_\textrm{outer}(h), \label{regularized in m}
\end{equation}
wherein
\begin{equation}
 \mathcal F_\textrm{outer}(h) = \ln\frac{2}{\sqrt{1-k^2}\mathrm{K}(k^2)}
 \label{Fouter}
\end{equation}
is a function of $h$ alone. Here, $\mathrm{K}$ is the complete elliptic integral of the first kind, where $k$ is determined as a function of $h$ from the implicit relation
\begin{equation}
\frac{\mathrm{K}(1-k^2)}{\mathrm{K}(k^2)}=h. \label{implicit}
\end{equation}
[We employ the convention where the argument of $\mathrm{K}(m)$ is the ``parameter'' --- the square of the ``modulus.'']

We note that in the limit $h\to\infty$ the solution of the problem \eqref{laplace w start}--\eqref{local} is well-known \citep{Brown:book}, with [cf.~\eqref{1/x}]
\begin{equation}
\pd{w^*}{y}(x,0) = \frac{1}{\sin(\pi x/2)}. \label{interior shear -1}
\end{equation}
It then follows from \eqref{regularized in m} that 
\begin{equation}
\mathcal F_\textrm{outer}(\infty) = \ln\frac{4}{\pi}.  
\label{large h}
\end{equation}
In the diametric limit, the solution of that problem is clearly $w^*\sim  1-y'/h$ whereby
\begin{equation}
\mathcal F_\textrm{outer} \sim \frac{\pi}{2h} \quad \text{as} \quad h\to0.
\label{small h}
\end{equation}

\begin{figure}[b]
\begin{center}
\includegraphics[scale=0.51,trim={1cm 1cm 0 0}]{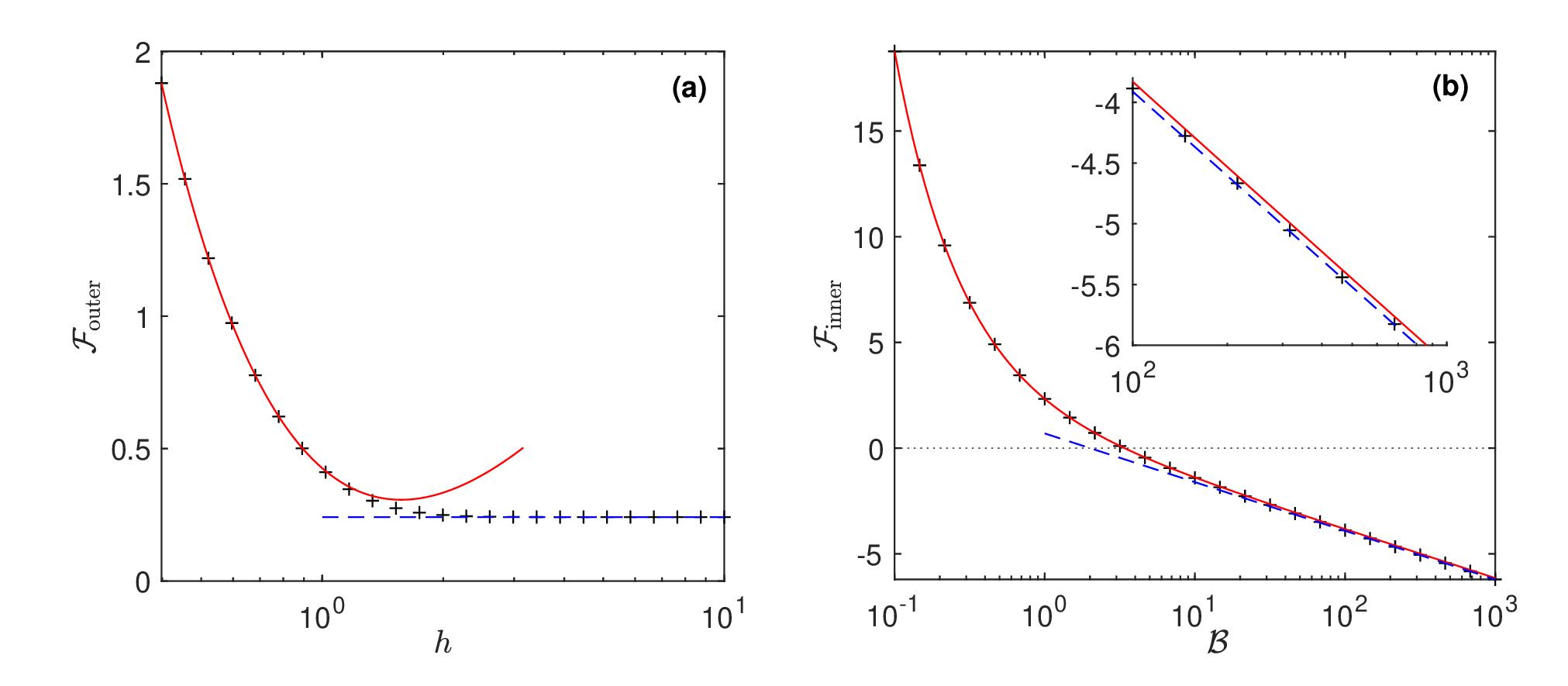}
\caption{The functions (a) $\mathcal{F}_{\textrm{outer}}(h)$ and (b) $\mathcal{F}_{\textrm{inner}}(\mathcal{B})$. Symbols: exact solutions obtained by numerical solution of \eqref{implicit} and \eqref{tau eqn} and evaluation of \eqref{Fouter} and \eqref{Finner}. Solid curves: $h\ll1$ approximation (\ref{F outer expansions}a) and $\mathcal{B}\ll1$ approximation (\ref{F inner expansions}a). Dashed curves: $h\gg1$ approximation (\ref{F outer expansions}b) and $\mathcal{B}\gg1$ approximation (\ref{F inner expansions}b). The inset demonstrates that the remarkable accuracy of the $\mathcal{B}\ll1$ approximation (\ref{F inner expansions}a) at moderate and large $\mathcal{B}$ is coincidental; indeed, note that the large $\mathcal{B}$ limit of that approximation disagrees with (\ref{F inner expansions}b), the true approximation in that limit, by an $\text{ord}(1)$ constant term, $1+\ln(\pi/8)\approx 0.065$, which just happens to be numerically small.}
\label{fig:Fs}
\end{center}
\end{figure}

The approximations \eqref{large h} and \eqref{small h} can also be inferred, and readily extended, from an asymptotic analysis of \eqref{Fouter} and \eqref{implicit} using the known expansions of $\mathrm{K}(m)$ for $m\searrow0$ and $m\nearrow1$ \citep{Abramowitz:book}. First, asymptotic inversion of \eqref{implicit} gives
\begin{subequations}
\begin{gather}
\ln(1-k) = -\frac{\pi}{h}+3\ln 2 +\mathrm{EST} \quad \text{as} \quad h\to0,\\
\ln k = -\frac{\pi}{2}h +2\ln 2 + \mathrm{EST} \quad \text{as} \quad h\to\infty,
\end{gather}  
\end{subequations}
where ``$\mathrm{EST}$'' stands for exponentially small terms. 
We then find from \eqref{Fouter},
\begin{subequations}
\begin{gather}
\mathcal{F}_{\mathrm{outer}}(h)=\frac{\pi}{2h}-\ln \frac{\pi}{h} + \mathrm{EST} \quad \text{as} \quad h\to0,\\
\mathcal{F}_{\mathrm{outer}}(h)=\ln\frac{4}{\pi}+\mathrm{EST} \quad \text{as} \quad h\to\infty.
\end{gather}
\label{F outer expansions}
\end{subequations}
These approximations are shown against a numerical evaluation of $\mathcal{F}_{\mathrm{outer}}$ in Fig.~\ref{fig:Fs}(a). 

\subsection{Inner region}
The inner region is analyzed by considering the distinguished sub-limit \eqref{dist thin}, this time holding fixed the stretched coordinates 
\refstepcounter{equation}
$$
X=x/\phi, \quad Y=-y/\phi.
\eqno{(\theequation{\mathit{a},\mathit{b}})}
$$
We here prefer to avoid using the symmetry about $X=0$, so that the geometry is as shown in Fig.~\ref{fig:inner}(a). We posit the leading-order inner approximation 
\begin{equation}
\tilde w^-(x,y;\phi) \sim W(X,Y)  \quad \text{as} \quad \phi\to0.
\label{def W}
\end{equation}

The leading-order inner problem consists of: (i) Laplace's equation,
\begin{equation}
\pd{^2W}{X^2}+\pd{^2W}{Y^2}=0;
\end{equation}
(ii) no slip at the top of the ridge,
\begin{equation}
\label{no-slip top}
W=0 \quad \text{at} \quad Y=\mathcal B \quad \text{for} \quad |X|<1;
\end{equation}
(iii) no slip at the sides of the ridge,
\begin{equation}
\label{no-slip sides}
W=0 \quad \text{at} \quad X = \pm1 \quad \text{for} \quad Y>\mathcal B;
\end{equation}
(iv) the Dirichlet condition
\begin{equation} \label{Dirichlet inner}
W=1 \quad \text{at} \quad Y=0;
\end{equation}
and (v) the matching condition [recall~\eqref{local}]
\begin{equation}
\label{W match}
W \sim \frac{2|\theta|}{\pi} \quad \text{as} \quad X^2+Y^2\to\infty.
\end{equation}
This problem is depicted (for $X>0$) in the inset of Fig.~\ref{fig:interior}. 

\subsection{Solution to the inner problem}
\begin{figure}[t!]
\begin{center}
\includegraphics[scale=0.4]{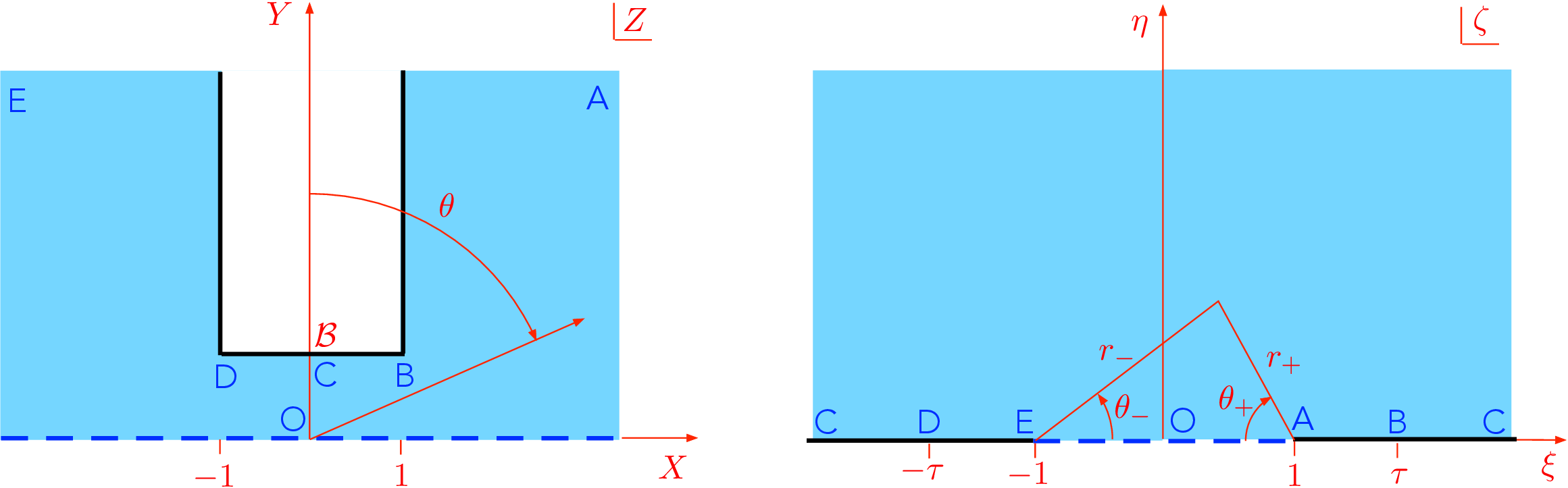}
\caption{(a) The inner region in the complex $Z$-plane. (b) Auxiliary $\zeta$-plane. Solid lines indicate a homogeneous Dirichlet condition, while  dashed lines indicate a unity Dirichlet condition.}
\label{fig:inner}
\end{center}
\end{figure}
Defining the complex variable $Z=X+\ii Y$ and an auxiliary complex variable $\zeta=\xi+\ii \eta$, we employ a conformal mapping $Z=f(\zeta)$ from the upper-half $\zeta$-plane to the fluid domain in the complex $Z$-plane. As shown in Fig.~\ref{fig:inner}, the map is chosen such that reflection across the $\eta$-axis corresponds to a reflection across the $Y$-axis, with $\zeta=0$ (point O) mapped to $Z=0$. Point A, $\zeta=1$, is mapped to infinity in the first quadrant of the $Z$-plane; point B, $\zeta=\tau$, is mapped to the ``corner'' $Z=1+\ii \mathcal B$; and the point at infinity, C, is mapped to $Z=\ii \mathcal B$, see Fig.~\ref{fig:inner}(b). 

On the $\xi$-axis, then,
condition \eqref{Dirichlet inner} gives
\begin{equation} \label{Dirichlet zeta}
W=1 \quad \text{for} \quad |\xi|<1,
\end{equation}
while conditions \eqref{no-slip top} and \eqref{no-slip sides} give 
\begin{equation}
W=0 \quad \text{for} \quad |\xi|>1. 
\end{equation}
Furthermore, the matching condition \eqref{W match} implies the presence of ``vortices'' at the critical points $\zeta=\pm1$,
\begin{equation} \label{vertices}
W \sim 1-\theta_\pm/\pi \quad \text{as} \quad \zeta\to\pm1,
\end{equation}
where the angles $\theta_\pm$ are indicated in Fig.~\ref{fig:inner}(b).

The harmonic function that satisfies \eqref{Dirichlet zeta}--\eqref{vertices}  is simply
\begin{equation}
W = 1 - \frac{\theta_+ + \theta_-}{\pi}.
\label{W real}
\end{equation}
It can be alternatively written
\begin{equation}
W = \operatorname{Im}\frac{\log(\zeta-1)-\log(\zeta+1)}{\pi}, 
\label{W complex}
\end{equation}
where
\stepcounter{equation}
$$
\log(\zeta-1) = \ln r_+ + \ii (\pi-\theta_+), \quad
\log(\zeta+1) = \ln r_- + \ii \theta_-.
\eqno{(\theequation{\mathit{a},\mathit{b}})}
$$

It remains to construct the mapping. We employ the Schwarz--Christoffel map \citep{Brown:book}
\begin{equation}\label{inner mapping}
f(\zeta) = -M\int_0^\zeta \frac{(\zeta'+\tau)^{1/2}(\zeta'-\tau)^{1/2}}{(\zeta'+1)^{3/2}(\zeta'-1)^{3/2}}\,\dd\zeta',
\end{equation}
where $M>0$ and $\tau>1$ are mapping parameters determined by two geometric requirements. The first, $f(\infty)=\ii\mathcal B$, gives
\begin{equation}
M \int_0^\infty \frac{(\eta^2+\tau^2)^{1/2}}{(\eta^2+1)^{3/2}} \, \dd\eta = \mathcal B \label{1}.
\end{equation}
The integral can be evaluated in terms of the complete elliptic integral of the second kind, $\mathrm{E}(m)$. This gives 
\begin{equation}
M\tau \mathrm{E}(1-\tau^{-2}) = \mathcal B \label{1}.
\end{equation}
[As for $\mathrm{K}(m)$, we employ the convention where the argument $m$ is the so-called parameter.] The second, $f(\infty)-f(\tau)=-1$, becomes
\begin{equation}
M \int_\tau^\infty \frac{(\xi^2-\tau^2)^{1/2}}{(\xi^2-1)^{3/2}} \, \dd\xi = 1 \label{2}.
\end{equation}
Again the integral can be evaluated in terms of elliptic integrals, giving 
\begin{equation}
M \tau \left[\mathrm{K}(\tau^{-2})-\mathrm{E}(\tau^{-2})\right] = 1 \label{2}.
\end{equation}
The parameter $\tau$ is accordingly determined from the implicit relation
\begin{equation}
\frac{\mathrm{E}(1-\tau^{-2})}{\mathrm{K}(\tau^{-2})-\mathrm{E}(\tau^{-2})}= \mathcal B. \label{tau eqn}
\end{equation}

The asymptotic behaviors of the functions $M(\mathcal{B})$ and $\tau(\mathcal{B})$ for small and large $\mathcal{B}$ are readily derived using the known asymptotic expansions of the elliptical integrals $\mathrm{E}(m)$ and $\mathrm{K}(m)$ as $m\searrow0$ and $m\nearrow1$ \citep{Abramowitz:book}. We thereby find 
\refstepcounter{equation}
\label{small B}
$$
M= \frac{2}{\pi}\mathcal{B}+\mathrm{EST}, \quad \ln(\tau-1) = -\frac{\pi}{\mathcal{B}}+\ln 8 -2 +\mathrm{EST} \quad \text{as} \quad \mathcal{B}\to0,
\eqno{(\theequation{\mathit{a},\mathit{b}})}
$$
and
\refstepcounter{equation}
\label{large B}
$$
M= \frac{2}{\pi^{1/2}}\mathcal{B}^{1/2}+O(\mathcal{B}^{-1/2}\ln\mathcal{B}), \quad \tau = \frac{\pi^{1/2}}{2}\mathcal{B}^{1/2}+O(\mathcal{B}^{-1/2}\ln\mathcal{B}) \quad \text{as} \quad \mathcal{B}\to\infty.
\eqno{(\theequation{\mathit{a},\mathit{b}})}
$$

\subsection{Shear force}
Consider now the drag integral \eqref{def cal F}.  We introduce the intermediate parameter
\begin{equation}
\phi\ll\delta\ll1,
\end{equation}
and decompose \eqref{def cal F} as
\begin{equation}
\mathcal{D} = \mathcal{D}_\textrm{inner} + \mathcal{D}_\textrm{outer},
\label{def split}
\end{equation}
in which we define the inner and outer contributions 
\refstepcounter{equation}
$$
\mathcal{D}_\textrm{inner}=\int_0^\delta\pd{\tilde w^-}{y}(x,0)\,\dd x, \quad
\mathcal{D}_\textrm{outer}=\int_\delta^1\pd{\tilde w^-}{y}(x,0)\,\dd x.
\eqno{(\theequation{\mathit{a},\mathit{b}})}
$$

Consider first the outer contribution. Using the leading-order outer velocity [cf.~\eqref{w- thin}], we approximate this contribution as 
\begin{equation}
\mathcal{D}_\textrm{outer} = \int_\delta^1\pd{w^*}{y}(x,0)\,\dd x.
\end{equation}
Motivated by \eqref{1/x}, we write 
\begin{equation}
\int_\delta^1\pd{w^*}{y}(x,0)\,\dd x=-\frac{2}{\pi}\ln\delta+\int_\delta^1\left[\pd{w^*}{y}(x,0)-\frac{2}{\pi x}\right]\,\dd x.
\label{subtract}
\end{equation}
Setting the lower integration limit $\delta$ to zero on the right-hand side of \eqref{subtract} incurs an algebraically small error in $\delta$. Using \eqref{Fouter}, we thus have 
\begin{equation}
\mathcal{D}_\textrm{outer}\sim -\frac{2}{\pi}\ln\delta + \frac{2}{\pi}\mathcal F_\textrm{outer}(h).
\label{S outer final really}
\end{equation}

Consider next the inner contribution. Using the leading-order inner velocity [cf.~\eqref{def W}], we approximate this contribution as
\begin{equation}
\label{inner}
\mathcal{D}_\textrm{inner}= -\int_0^{\delta/\phi}\pd{W}{Y}(X,0)\,\dd X.
\end{equation}
Since $Z=\delta/\phi$ is $\gg1$, we can write the pre-image of this point as $\zeta=1-\epsilon$, where $\epsilon$ is a positive small parameter defined by the relation [cf.~\eqref{inner mapping}]
\begin{equation}
f(1-\epsilon) = \delta/\phi. \label{def eps}
\end{equation}
With this definition, \eqref{inner} becomes 
\begin{equation}
\label{inner in zeta}
\mathcal{D}_\textrm{inner}= -\int_0^{1-\epsilon}\pd{W}{\eta}(\xi,0)\,\dd \xi.
\end{equation}
From \eqref{W real}, we obtain 
\begin{equation}
\pd{W}{\eta}(\xi,0) = -\frac{1}{\pi}\left(\frac{1}{1-\xi}+\frac{1}{1+\xi}\right),
\end{equation}
so that \eqref{inner in zeta} gives 
\begin{equation} \label{S inner in eps}
\mathcal{D}_\textrm{inner} \sim \frac{1}{\pi}\ln\frac{2}{\epsilon}.
\end{equation}

We need to transform \eqref{S inner in eps} into a form where the dependence upon $\delta/\phi$ is explicit. Since $\delta/\phi\gg1$, it suffices to asymptotically invert \eqref{def eps} to obtain an approximation of $\epsilon$ in that limit. To this end, we need the expansion of [cf.~\eqref{inner mapping}] 
\begin{equation}
f(1-\epsilon)=M\int_0^{1-\epsilon}\frac{(\tau^2-\xi^2)^{1/2}}{(1-\xi^2)^{3/2}}\,d\xi,
\end{equation}
as $\epsilon\to0$. In that limit, the dominant contribution to the integral is from the vicinity of the upper boundary, so that to leading order the integrand can be approximated by the power $(\tau^2-1)^{1/2}/[2(1-\xi)]^{3/2}$; integration then yields the leading-order approximation 
\begin{equation}
f(1-\epsilon)\sim \frac{M(\tau^2-1)^{1/2}}{2^{1/2}\epsilon^{1/2}} \quad \text{as} \quad \epsilon\to0. 
\end{equation}
It follows from \eqref{def eps} that 
\begin{equation}
\epsilon \sim \frac{M^2(\tau^2-1)}{2(\delta/\phi)^2} \quad \text{as} \quad \delta/\phi\to+\infty. 
\label{eps invert}
\end{equation}

Using \eqref{eps invert} in \eqref{S inner in eps}, we find 
\begin{equation}
\mathcal{D}_\textrm{inner} \sim 
\frac{2}{\pi}\ln\frac{\delta}{\phi}+\frac{2}{\pi}\mathcal F_\textrm{inner}(\mathcal B),
\label{S inner final}
\end{equation}
wherein
\begin{equation}
\mathcal F_\textrm{inner}(\mathcal B) = \ln\frac{2}{M\sqrt{\tau^2-1}}.
\label{Finner}
\end{equation}

Asymptotic expansions of $\mathcal{F}_\textrm{inner}(\mathcal{B})$ for small and large $\mathcal{B}$ are readily derived by substituting expansions \eqref{small B} and \eqref{large B} into \eqref{Finner}: 
\begin{subequations}
\begin{gather}
\mathcal{F}_\textrm{inner}(\mathcal{B})=\frac{\pi}{2\mathcal{B}}+\ln\frac{\pi}{4\mathcal{B}}+1+\mathrm{EST} \quad \text{as} \quad \mathcal{B}\to0,\\
\mathcal{F}_{\textrm{inner}}(\mathcal{B})=\ln\frac{2}{\mathcal{B}}+O(\mathcal{B}^{-1}\ln\mathcal{B}) \quad \text{as} \quad \mathcal{B}\to\infty. 
\end{gather}
\label{F inner expansions}
\end{subequations}
These approximations are shown against a numerical evaluation of $\mathcal{F}_{\mathrm{inner}}$ in Fig.~\ref{fig:Fs}(b). 

\subsection{Slip length}
Upon adding the outer contribution \eqref{S outer final really} and the inner contribution \eqref{S inner final}, the dependence upon $\delta$ disappears, as it must. Thus, \eqref{def split} yields the drag as 
\begin{equation}
\mathcal{D} \sim
\frac{2}{\pi} \left[ \ln(1/\phi)+ \mathcal F_\textrm{outer}(h) + \mathcal F_\textrm{inner}(\mathcal B) \right].
\end{equation}
The scaled slip length then follows from \eqref{slip is 1/D}, 
\begin{equation}
\tilde\lambda
\sim \frac{\pi}{2 \left[\ln (1/\phi)+ \mathcal F_\textrm{outer}(h) + \mathcal F_\textrm{inner}(\mathcal B) \right]},
 \label{lambda -1}
\end{equation}
with algebraic accuracy in the distinguished sub-limit \eqref{dist thin}. 

It is instructive to consider how the result \eqref{lambda -1} simplifies for small and large $\mathcal{B}$. For ``ultra-thin'' ridges, $\mathcal{B}\gg1$, we use (\ref{F inner expansions}b) and the definition $\mathcal{B}=b/\phi$ to find
\begin{equation}
\tilde\lambda \sim  \frac{\pi}{2 \left[ \ln({2}/{b})+\mathcal F_\textrm{outer}(h)  \right]}. 
 \label{lambda -1 thin}
\end{equation}
The significance of this approximation will be discussed in the following section. 
The other extreme, $\mathcal{B}\ll1$, corresponds to relatively thick ridges. Using the leading order of (\ref{F inner expansions}a), we find 
\begin{equation}
\tilde\lambda
\sim \mathcal B. 
 \label{lambda -1 thick}
\end{equation}
This limiting behavior is nothing but the algebraic approximation \eqref{algebraic regime primitive}, see \eqref{mu -1 scaling} and \eqref{def B}.

We can also examine the asymptotic behavior of \eqref{lambda -1} for small and large $h$. The expansion (\ref{F outer expansions}b) shows that $\mathcal{F}_{\mathrm{inner}}(h)$ approaches a constant as $h\to\infty$, rendering that limit regular. In contrast, in the shallow-cavity limit $h\ll1$, $\mathcal{F}_{\mathrm{inner}}(h)$ is singular; using (\ref{F outer expansions}a), we find that \eqref{lambda -1} reduces to $\tilde{\lambda}\sim h$. Expressing this approximation as $\lambda \sim \mu^{-1}h$ reveals that the interior problem becomes invalid for ultra-shallow grooves,    $h=O(\mu)$. We do not consider this extreme scenario. 

\section{Discussion} \label{Sec:Discussion}
To summarize our results, we refer back to the ``phase map'' in Fig.~\ref{fig:Phase}, which delineates the asymptotic regimes and corresponding approximations in the $(b,\phi)$ parameter space for $\mu\ll1$. In this discussion, we assume that $h$ is held fixed as $\mu\to0$, and is thus of order unity in that limit. 

Our analysis is based on two key limits, marked by the filled circles in Fig.~\ref{fig:Phase}. In key limit I, where $\mu\to0$ with $b$ and $\phi$ held fixed, we find $\lambda\sim \mu^{-1}\tilde{\lambda}(b,\phi)$, where $\tilde{\lambda}(b,\phi)$ is governed by the interior problem (see Sec.~\ref{sec:nearly}). In key limit II, where $\mu\to0$ with $b=\text{ord}(\mu)$ and $\phi$ held fixed, we find $\lambda\sim \Lambda(b/\mu,\phi)$, where $\Lambda(b/\mu,\phi)$ is governed by the (exterior) generalized Philip problem (see subsection \ref{ssec:generalized}). As shown in Fig.~\ref{fig:Phase}, the validity of these two approximations extends to \emph{overlapping} asymptotic regions in the $(b,\phi)$ plane beyond the key limits in which they have been derived. In these extended domains, but away from the key limits, asymptotic analysis of the interior and generalized Philip problems furnish approximations of the rescaled slip lengths $\tilde{\lambda}(b,\phi)$ and $\Lambda(b/\mu,\phi)$, respectively. 

Our interest is in the slight-encapsulation regime, $b\ll1$, considering both moderate and small $\phi$. Referring to Fig.~\ref{fig:Phase}, we see that this regime is well described by the generalized Philip problem together with the small-$b$ approximations to the interior problem; as discussed at the end of this section, this union still excludes an exotic region where both $b$ and $\phi$ are exponentially small in $\mu$. Accordingly, we have fully solved the generalized Philip problem (as well as analyzed its limiting cases), whereas for the interior problem we focused on small-$b$ approximations rather than a general solution. 

We now describe the phase map in more detail, starting with the scenario where $\phi$ is held fixed while $b$ is reduced from order unity. The starting point is thus key limit I, where the slip length is huge, $\text{ord}(\mu^{-1})$. Once $b$ becomes small, we enter the algebraic regime, where $\lambda\sim b/(\mu\phi)$; the slip-length is diminished, yet remains algebraically large for $b\gg\mu$. We reach key limit II when $b$ becomes comparable to $\mu$; the slip length is then of order unity, provided by our numerical solution to the generalized Philip problem (see Sec.~\ref{ssec:numericalofNavier}). When $b$ is further reduced, the generalized Philip problem continues to hold but simplifies to Philip's classical problem corresponding to a superhydrophobic surface, with $\phi$ becoming the genuine solid fraction. Thus, the slip length remains $\text{ord}(1)$, and is provided in closed form by \eqref{Philip}. 

The above picture extends to small $\phi$, with the transition from a huge $\text{ord}(\mu^{-1})$ slip length to the algebraic approximation $\lambda\sim b/(\mu\phi)$ occurring at $\phi=\text{ord}(b)$ --- a distinguished sub-limit of the interior problem in which $\Lambda$ is given by \eqref{slip in Lambda} --- and the transition from the algebraic regime to Philip's classical solution occurring at $\phi=\text{ord}(b/\mu)$ --- a distinguished sub-limit of the generalized Philip problem in which $\tilde{\lambda}$ is given by \eqref{lambda -1}. For small $\phi$, however, Philip's expression for the slip length \eqref{Philip} grows from $\text{ord}(1)$ to $\text{ord}(\ln\phi)$, simplifying to  \eqref{Philip log}, and the slip length in the huge-slip regime slightly diminishes from $\text{ord}(\mu^{-1})$ to $\text{ord}(\mu^{-1}\ln^{-1}b)$, simplifying to \eqref{lambda -1 thin}.   

The different regimes in Fig.~\ref{fig:Phase} can also be interpreted in terms of the integral relations \eqref{int out} and \eqref{int in}, which represent the balance between the drag on the flat fluid interface and the imposed far-field stress. In the regime of the generalized Philip problem, the drag is always dominated by the fluid interface adjacent to the ridge top. Within that regime, the transition from Philip's classical solution to the algebraic regime represents the crossover from solid-fraction-limited resistance to encapsulation-limited resistance, respectively, with both solid fraction and encapsulation being important in the distinguished sub-limit $\phi=\text{ord}(b/\mu)$. In the interior-problem regime, the drag is generally supported by the whole interface. In the distinguished sub-limit $b=\text{ord}(\phi)$, the inner region near the ridge top, and the outer region corresponding to the cavity scale, both contribute to the drag at leading algebraic order, with the overlap between these regions resulting in a logarithmically enhanced drag. On the $b\gg\phi$  side of this distinguished scaling, the outer region dominates the drag, whereas on the $b\ll\phi$ side, the inner region does. 

As alluded to earlier, we anticipate that both the interior and generalized Philip problem lose validity when $b$ and $\phi$ are both exponentially small in $\mu$. Indeed, the interior problem hinges upon the assumption of an asymptotically large and approximately uniform exterior flow. However, in the distinguished sub-limit of that problem, $\phi\ll1$ with $b/\phi$ fixed, the flow (which scales as the slip length) diminishes to $\text{ord}(1)$ as $\mu \ln \phi$ grows to $\text{ord}(1)$, see \eqref{lambda -1}; a similar conclusion is obtained from the $\phi\ll b\ll1$ approximation of that problem, in which case a contradiction is seen when $\mu \ln b$ grows to $\text{ord}(1)$, see \eqref{lambda -1 thin}. Similarly, the generalized Philip problem assumes that the shear stress on the cavity scale can be neglected. Accordingly, the $\text{ord}(\ln \phi)$ scaling of the velocity in the distinguished sub-limit $\phi=\text{ord}(b/\mu)$ suggests a contradiction when $\mu\ln\phi$ grows to $\text{ord}(1)$. 

A special case of the above exotic exponential regime is that where $b=\phi=0$. 
As already discussed in the Introduction, that scenario was recently analyzed by \citet{Peng:25}. They found  $\lambda=\text{ord}(\mu^{-1/2})$, with the dominant drag contribution arising from an exponentially small tip region where $\mu\ln\rho=\text{ord}(1)$, $\rho$ here being the dimensionless distance from the tip. It is plausible that, in our setup, a similar phenomenon occurs when $\mu\ln \phi$ and $\mu\ln b$ are $O(1)$ --- defining the exponential regime shown in Fig.~\ref{fig:Phase} --- with the slip-length scaling coinciding with that identified by \citet{Peng:25}. The generalization of \citet{Peng:25} to include the effects of exponentially small $b$ and $\phi$ is left open. 

\bibliography{refs_new}
\end{document}